\documentclass[twocolumn,floatfix]{revtex4}%
\usepackage[dvipdfmx]{graphicx}%
\usepackage{amsmath}%
\usepackage{amsfonts}
\usepackage{amssymb}
\usepackage{hhline}
\usepackage{bm}
\usepackage{mathrsfs}
\usepackage{color}
\def\d{{\partial}}
\def\s{{\sigma}}
\def\e{{\epsilon}}
\def\k{{ {\bm k} }}

\def\q{{ {\bm q} }}
\def\Q{{ {\bm Q} }}

\def\0{{ {\bm 0} }}
\def\w{{\omega}}
\def\a{{\alpha}}

\def\dd{{\delta}}

\allowdisplaybreaks[4]

\begin{document}
\title{
A Rigorous Formalism of Unconventional Symmetry Breaking in Fermi Liquid Theory\\
and Its Application to Nematicity in FeSe
}
\author{
Rina Tazai, Shun Matsubara, Youichi Yamakawa, Seiichiro Onari, and Hiroshi Kontani}

\date{\today }

\begin{abstract}
Unconventional symmetry breaking due to nonlocal order parameters
has attracted considerable attention in many strongly correlated metals.
Famous examples are the nematic order in Fe-based superconductors 
and the star-of-David charge density order in kagome metals.
Such exotic symmetry breaking in metals is a central issue of 
modern condensed matter physics, while its theoretical foundation
is still unclear in comparison with the well-established 
theory of superconductivity.
To overcome this difficulty, here we introduce the 
``form factor'' that generalizes the nonlocal order parameter
into the Luttinger-Ward (LW) Fermi liquid theory.
We then construct a rigorous formalism of the ``density-wave equation''
that gives the thermodynamically stable form factor,
similarly to the superconducting-gap equation.
In addition, a rigorous expression of the 
Ginzburg-Landau free-energy for the unconventional order is presented
to calculate various thermodynamic properties.
In the next stage, we apply the derived formalism to a typical Fe-based superconductor FeSe,
by using the one-loop LW function 
that represents the free-energy gain due to the interference among paramagnons.
The following key experiments are naturally explained:
(i) Lifshitz transition (=disappearance of an electron-pocket)
due to the bond+orbital order below $T_{\rm c}$.
(ii) Curie-Weiss behavior of the nematic susceptibility 
at higher $T$, and the deviation from the Curie-Weiss behavior 
at lower $T$ near the nematic quantum-critical-point.
(iii) Scaling relation of the specific heat jump at $T_{\rm c}$,
$\Delta C/T_{\rm c} \propto T_{\rm c}^b$ with $b\sim3$. 
(Note that $b=0$ in the BCS theory.)
These results lead to a conclusion that 
the nematicity in FeSe is the bond+orbital order
due to the ``paramagnon interference mechanism''.
The present theory paves the way for solving
various unconventional phase transition systems.

\end{abstract}

\address{
Department of Physics, Nagoya University,
Furo-cho, Nagoya 464-8602, Japan. 
}

\sloppy

\maketitle

\section{Introduction}
\label{sec:Intro}

Recently, rich symmetry breaking phenomena due to unconventional 
order parameters have attracted considerable attention
in various electron systems.
Famous examples are the $C_2$-symmetric nematic order 
in various Fe-based superconductors
\cite{Chubukov-nematic-rev,Yamakawa-PRX2016,Onari-FeSe,Fernandes-Nature}.
It has been established that the nematic state is driven by
the electron-correlation, thanks to the 
``electronic nematic susceptibility'' measurements
performed by the shear-modulus analysis
\cite{Yoshizawa,Fernandes1,Bohmer},
the Raman spectroscopy
\cite{Gallais,Gallais2,Raman-spectroscopy},
and the elastoresistivity measurements
\cite{Fisher-review,Hosoi,B2g-Ishida,Fisher-Science2016}.
Similar nematic orders are also observed in 
magic-angle-twisted-bilayer-graphene
\cite{Fernandes-TBG,Onari-TBG},
titanium pnictide oxide 
\cite{Nakaoka-Ti},
and cuprate superconductors
\cite{Sato-CDW,Murayama-CDW,Kawaguchi-CDW,Tsuchiizu-CDW}.
Also, correlation-driven density-wave (DW)  with nonzero wavevector
($\q\ne{\bm0}$) has attracted increasing attention recently.
Famous examples are the star-of-David DW state 
in kagome metal CsV$_3$Sb$_5$
\cite{Tazai-kagome,kagome-Thomale2013,kagome-SMFRG,kagome-Balents2021}
the CDW states in cuprate superconductors
\cite{Yamakawa-CDW,Kawaguchi-CDW,Tsuchiizu-CDW,Sachdev-CDW1,Husemann,Kivelson-spin-nematic,DHLee-spin-nematic},
and the multopole DW states in heavy fermion systems
\cite{Tazai-CeB6,Tazai-CeCu2Si2-1,Tazai-CeCu2Si2-2}.
Furthermore, more exotic odd-parity DW orders that accompany the charge 
or spin loop current have been discovered
in kagome metals
\cite{mSR_TRS,Kerr},
iridates
\cite{Murayama},
and cuprates
\cite{Varma}.

We call these DW states ``unconventional'' because 
they have non-local and non-$s$-wave order parameters, 
in analogy to unconventional (non-$s$-wave) superconductivity.
\color{black}
For example, the order parameter of the $d$-wave bond order is
$O_{i,j}=\bar{O}(\delta_{|x-x'|,1}\delta_{y-y',0}-\delta_{x-x',0}\delta_{|y-y'|,1})$,
where $(x,y)$ and $(x',y')$ are the integer coordinates of $i$ and $j$ sites
\cite{Nersesyan,Kivelson-NJP,Halboth:2000tt,Davis:2013ce,Yamakawa-CDW,Kawaguchi-CDW,Tsuchiizu-CDW}.
(In high contrast, the conventional magnetic order
$m_i = n_{i\uparrow}-n_{i\downarrow}$ is local.)
However, simple local spin-density-wave (SDW) is inevitably derived 
within the mean-field-level approximations
for the Hubbard models with screened Coulomb interactions
\cite{Tazai-JPSJ-fRG,Kontani-sLC}.
Therefore, beyond-mean-field many-electron theories are necessary
to understand the unconventional DW states.
This is a difficult but very interesting theoretical problem,
and this is a central issue of modern condensed matter physics.
On the other hand, these unconventional DW states we are interested in
are metallic, so the itinerant picture will be fruitful.
In addition, the $U(1)$ gauge symmetry is preserved.
Thus, it is promising to construct the formalism of the DW states
on the basis of the microscopic Fermi liquid theory
\cite{AGD,LW,Baym-Kadanoff,Baym}.

In general, the DW state at wavevector $\q$
originates from the particle-hole (p-h) pairing condensation,
$D_\k^{\q\s}= (1-P_{0})\langle c_{\k+\q,\s}^\dagger c_{\k,\s}\rangle$,
where $c_{\k,\s}$ is the electron annihilation operator,
$\k$ is the momentum, and $\s \ (=\pm1)$ is the spin index
\cite{Nersesyan,Tazai-JPSJ-fRG,Kontani-sLC}.
$P_{0}$ represents the projection onto the totally-symmetric state
with respect to the space-group and the time-reversal
\cite{Tazai-JPSJ-fRG,Kontani-sLC}.
Rich classes of the DW states are determined by the 
symmetry of the p-h condensation $D_\k^{\q\s}$.
For example, a simple spin-DW state is given as
the $\k$-independent function $D_\k^{\q\s}=m\s$.
The realized DW state $D_\k^{\q\s}$ should be uniquely determined 
as the extremum of the free energy.

In Fermi liquids, the single-electron kinetic term 
between sites $i$ and $j$ is given by 
$t_{i,j}^{\s}=t_{i-j}^0+\Sigma_{i,j}^\s$,
where $t_{i-j}^0$ is the hopping integral of the bare Hamiltonian,
and $\Sigma_{i,j}^\s$ is the self-energy due to the correlation
between other electrons.
Here, we define the symmetry breaking part of the self-energy
\cite{Tazai-JPSJ-fRG,Kontani-sLC}:
\begin{eqnarray}
\delta t_{i,j}^\s \equiv (1-P_{0})\Sigma_{i,j}^\s.
\end{eqnarray}
%
In the absence of the DW order ($T\ge T_{\rm c}$),
we obtain $\delta t_{i,j}^\s=0$ by definition.
When the DW order emerges ($T\le T_{\rm c}$),
$\delta t_{i,j}^\s$ becomes finite due to nonzero $D_\k^{\q\s}$.
Thus, $\delta t_{i,j}^\s$ is the energy-dimension
order parameter of the DW state.

Here, we consider the DW state at a constant wavevector $\q$.
For convenience, 
we introduce the ``form factor $\delta t^{\q\s}_{\k}$'' 
that is the Fourier transform of $\delta t_{i,j}^\s$
\cite{Tazai-JPSJ-fRG,Kontani-sLC}:
\begin{eqnarray}
\delta t^{\q\s}_{\k} \equiv \frac1N \sum_{ij} \delta t_{ij}^\s e^{-i\k\cdot(\bm{r}_i-\bm{r}_j)} e^{-i\q\cdot \bm{r}_i} ,
\label{eq:formfactor1}
\end{eqnarray}
where $\bm{r}_i$ is the position of site $i$.
The classification of the symmetry of the form factor 
is presented in Sect. \ref{sec:formfactor}.

The theoretical way to derive the form factor $\delta t^{\q\s}_{\k}$ 
has not been established yet.
The aim of the present study is to establish an exact framework
to derive $\delta t^{\q\s}_{\k}$,
based on which we can construct reliable and useful 
approximate theories.
In the statistical mechanics,
the symmetry breaking with $\delta t^{\q\s}_{\k}$ occurs
when the grand potential $\Omega$ is reduced.
In other words,
$\delta t^{\q\s}_{\k}$ is uniquely determined as the stationary state 
with the minimum grand potential.
In strongly correlated Fermi liquids,
a rigorous formalism of the grand potential $\Omega$
is given by the Luttinger-Ward (LW) theory
\cite{LW},
In the LW theory, the LW function $\Phi[G]$, 
which is the functional of electron Green function $G$, plays a central role.
The self-energy $\Sigma$ and the irreducible two-particle interaction $I$
are uniquely derived from the functional derivatives of $\Phi[G]$
\cite{LW}.

In this paper, we first introduced the 
``form factor $\delta t^{\q\s}_{\k}$''
into the LW theory to analyze the symmetry breaking at wavevector $\q$.
Its $(\k-\q/2)$-dependence represents the symmetry of the DW. 
We next derived the ``DW equation'' to obtain the form factor 
that minimize the LW ground potential below $T_{c}$.
\color{black}
In this theory, 
the optimized $\delta t^{\q\s}_{\k}$ is uniquely obtained 
because the DW equation is
equivalent to the thermodynamic stationary condition.
This formalism enables us to study rich variety of
electron-correlation-driven DW states ($\delta t^{\q\s}_{\k}$)
without assuming any symmetry,
like the analysis of the superconducting (SC) states 
($\Delta_\k^{\s\s'}$) based on the SC gap equation.
In addition, we derive an exact expression of the 
Ginzburg-Landau (GL) free energy, $F\propto a_\q\phi^2$,
where $\phi$ is the amplitude of the DW order.
The coefficient $a_\q$
is uniquely related to the eigenvalue of the DW equation $\lambda_\q$.
Thus, we can calculate the
thermodynamic properties and the stability of the DW state.

In the next stage, we apply the derived DW equation to FeSe
by using the one-loop LW function $\Phi_{\rm FLEX}[G]$
that represents the quantum interference among paramagnons.
This theory naturally explains the following 
essential experimental reports:
{\bf (i)} Nematic Fermi surface (FS) and the Lifshitz transition
due to bond+orbital order
\cite{FeSe-ARPES-Suzuki,FeSe-Lif1,FeSe-Lif2,Eremin}.
{\bf (ii)} Curie-Weiss (CW) behavior of the nematic susceptibility $\chi_{\rm nem}$
at higher temperatures,
and the deviation from the CW behavior at lower temperatures
near the nematic quantum-critical-point (QCP).
\cite{Hosoi,B2g-Ishida,Terashima,Fisher-Science2016}.
{\bf (iii)} Scaling relation of the specific heat jump at $T_{\rm c}$,
$\Delta C/T_{\rm c} \propto T_{\rm c}^b$ ($b\sim3$).
This relation naturally explains the 
smallness of (or undetected) $\Delta C/T_c$ 
reported in several nematic systems, 
such as RbFe$_2$As$_2$ ($T_c\sim40$K) \cite{Mizukami-B2g}
and cuprate superconductors ($T_c\sim200$K) 
\cite{Sato-CDW,Murayama-CDW}. 
In cuprates, the nematic transition occurs at 
the pseudogap temperature $T^*$ 
\cite{Sato-CDW,Murayama-CDW}, 
while no anomaly in specific heat at $T=T^*$ 
has been reported previously.
\color{black}

Interestingly, recent experiments have revealed that the 
nematic QCP is clearly separated from the magnetic QCP
in Fe(Se,S) and Fe(Se,Te) \cite{Shibauchi-nemQCP}
in addition to Na(Fe,Co)As \cite{Zheng-NaFeAs}.
Such clear separation between two QCPs
in addition to the key points {\bf (i)}-{\bf (iii)}
are naturally understood by the present theory.
Therefore, the nematicity in FeSe is the bond+orbital order
due to the ``paramagnon interference mechanism''
\cite{Onari-SCVC,Yamakawa-PRX2016,Onari-FeSe,Tazai-JPSJ-fRG}.

Here, we construct the DW equation based on the LW theory.
The LW theory is expected to be valid for various strongly-correlated 
metals except for the vicinity of the localized Mott states.
Thus, the present theory is expected to pave the path to understanding 
the useful concept of the ``vestigial nematic order'' 
from the itinerant picture.

\subsection{Form Factor}
\label{sec:formfactor}

Here, we discuss the rich variety of unconventional DW states
\cite{Schultz,Affleck,Nersesyan,Tazai-JPSJ-fRG,Kontani-sLC}
by classifying the symmetry of the form factor.
The exotic DW states are represented by the non-local order parameter
$\delta t_{ij}^{\sigma}$, which is parameterized by different site indices $i$ and $j$.
Then, the effective hopping integral is 
$t_{i,j}^\s=t_{i-j}^0+\delta t_{ij}^{\sigma}$,
where $t_{i-j}^0$ is the original hopping integral with $A_{1g}$ symmetry.
In conventional charge (spin) orders, the order parameter is expressed as 
$\delta t_{ij}^{\uparrow}= + (-)\delta t_{ij}^{\downarrow}$
with $i=j$, respectively.

\begin{figure*}[htb]
\includegraphics[width=.8\linewidth]{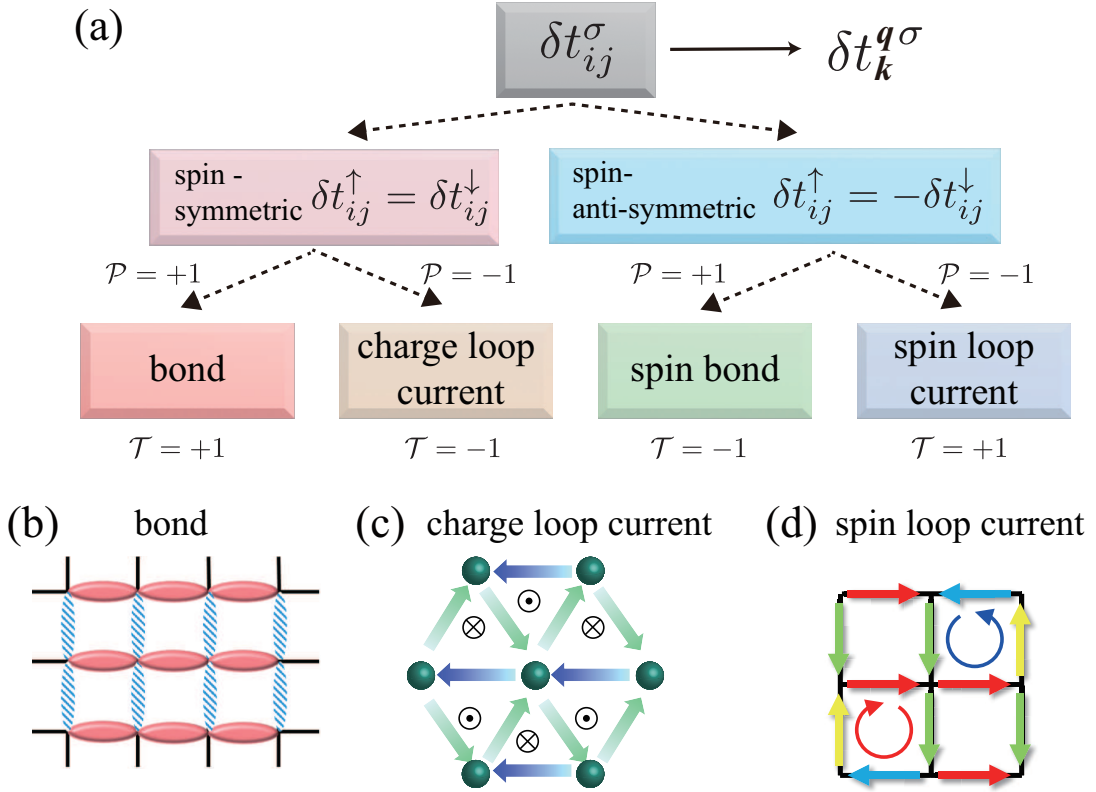}
\caption{
(a) Classification of nonlocal order parameters 
(quantum liquid crystal states).
Spin-symmetric (-antisymmetric)
order corresponds to the charge (spin) channel order.
The four states are labeled as $({\cal P},{\cal T})$,
where ${\cal P}=\pm1$ (${\cal T}=\pm1$) 
is the parity of the inversion-symmetry (time-reversal-symmetry).
(b) $d$-wave bond order with $({\cal P},{\cal T})=(+1,+1)$.
(c) Charge-loop-current (cLC) order with $({\cal P},{\cal T})=(-1,-1)$
in Ref. \cite{Tazai-cLC}.
(d) Spin-loop-current (sLC) order with $({\cal P},{\cal T})=(-1,+1)$
in Ref. \cite{Kontani-sLC}.
}
\label{fig:model}
\end{figure*}

In recent years, in contrast, unconventional 
non-local orders given by $\delta t_{ij}^{\sigma}$ with $i\neq j$
have been discovered and attracted increasing attentions.
Here, we assume the Hermitian order parameter
\cite{Tazai-JPSJ-fRG,Kontani-sLC}:
\begin{eqnarray}
&&\delta t_{ij}^\s = (\delta t_{ji}^\s)^* ,
\\
&&\delta t^{\q\s}_{\k}=(\delta t^{-\q\s}_{\k+\q})^* ,
\label{eq:Hermite}
\end{eqnarray}
$\delta t_{ij}^{\sigma}$ is classified into four channels
in terms of parity symmetry (${\cal P}=\pm1$)
and time-reversal symmetry (${\cal T}=\pm1$)
as shown in Fig.\ref{fig:model} (a).
Below, we discuss the case of $\q={\bm0}$.
First, we consider the case of $\delta t_{ij}^{\uparrow}=\delta t_{ij}^{\downarrow}$.
When $\delta t_{ij}$ is real, 
the bond order with $({\cal P},{\cal T})=(+1,+1)$ is realized.
As an example, the $d$-wave bond order in square lattice model
is shown in Fig. \ref{fig:model} (b). 
When $\delta t_{ij}$ is pure imaginary,
the charge-loop current (cLC) with $({\cal P},{\cal T})=(-1,-1)$ is realized.
Its form factor in $\k$-space is $\delta t_{\k} \propto \cos k_x - \cos k_y$.
An example of the cLC order in anisotropic triangular lattice model
is shown in Fig. \ref{fig:model} (c).

Next, we consider the case of $\delta t_{ij}^{\uparrow}=-\delta t_{ij}^{\downarrow}$.
When $\delta t_{ij}$ is real, 
the spin-bond order with $({\cal P},{\cal T})=(+1,-1)$ appears.
When $\delta t_{ij}$ is pure imaginary,
the spin-loop current (sLC) order with $({\cal P},{\cal T})=(-1,+1)$ appears.
An example of the sLC order is shown in Fig. \ref{fig:model} (d).

In addition, the translational symmetry is violated
when the DW wavevector $\q$ is nonzero.
Furthermore, orbital orders
\cite{Onari-SCVC}, 
valley orders
\cite{Onari-TBG},
and multipole orders
\cite{Tazai-CeB6}
emerge when the Wanner functions
possess multiple degrees of freedom.
These rich non-local DW states are called the quantum-liquid-crystal (QLC) order
\cite{Tazai-JPSJ-fRG},
We note that, in a simple single-site model,
the cLC and sLC orders in real space ($\delta t_{ij}^\s=-\delta t_{ij}^\s$) 
are pure imaginary according to the Hermitian condition.
However, the Fourier transformation of such current order
$\delta t^{\q\s}_{\k}$ becomes real.

\subsection{Stoner Theory for Ferro-Magnetic Transition}
\label{sec:stoner}

Here, we briefly review the mean-field theory of the 
ferro-magnetic (FM) transition and derive a simplified DW equation 
and GL \rm{free} energy. 
This explanation will be useful to understand the 
derivation of the DW equation based on the LW theory in later sections.
It is also understood that the 
mean-field theory is insufficient to explain the nematic order.
\color{black}
Below, we consider the following Hubbard model,
\begin{eqnarray}
H&=&\sum_{i\neq j, \sigma}t_{i-j}^0c^{\dagger}_{i \sigma}c_{j\sigma}+H_{\rm{I}} ,
\\
H_{\rm{I}}&=&\sum_{i\sigma } U c^{\dagger}_{i\sigma}c_{i\sigma} c^{\dagger}_{i\bar{\sigma}}c_{i\bar{\sigma}}=\frac{U}{4} \sum_i (n_i^2-m_i^2), 
\label{eq:1}
\end{eqnarray}
%
where $n_i=n_{i\uparrow}+n_{i\downarrow}$ and $m_i=n_{i\uparrow}-n_{i\downarrow}$.
The conduction electron energy is 
$\e_\k = \frac1N \sum_{i,j}t_{i-j}^0 e^{i(\bm{r}_i-\bm{r}_j)\cdot\k}$.
In the mean-field approximation, 
the magnetic order is expressed as the spin-dependent $\dd t$: 
$\delta t_{ii}^{\uparrow}=-\delta t_{ii}^{\downarrow}=-\frac{U}{2}m_i$. 
Here, we set $M_i \equiv U m_i/2$.
In the case of the FM ($q=0$) order, $M_i=M$.
Thus, the grand potential is given by
\begin{eqnarray}
\Omega_{\rm{MF}}&=& \frac{T}{N} \sum_{k,\sigma} \ln G_{k\sigma}^{\rm{MF}}+\frac{1}{U} M^2 \nonumber \\
&=& - \frac{1}{N} \sum_{\k,\sigma}  \ln (1+e^{-\beta (\epsilon_{\k\s}-\mu)})+\frac{1}{U}M^2,
\label{eq:stoner1}
\end{eqnarray}
where $k\equiv (\k,\e_n=(2n+1)\pi T)$,
$\epsilon_{\k\sigma}=\epsilon_{\k}+M\sigma$,
and $G_{k\sigma}^{\rm{MF}}=(i\epsilon_n-\epsilon_{\k\s}+\mu)^{-1}$
is the electron Green function.
From the stationary condition of the $\Omega_{\rm{MF}}$, which is 
given by ${\d \Omega_{\rm{MF}}}/{\d M}=0$, 
the mean-field equation for $M$ is obtained as
%
\begin{eqnarray}
M=-\frac{U}{2N} \sum_{\k\sigma} f(\epsilon_{\k\sigma}) \sigma .
\label{eq:stoner2}
\end{eqnarray}
Next, we derive the linearized mean-field equation 
by linearizing the right-hand side of Eq. (\ref{eq:stoner2}).
It is given as
\begin{eqnarray}
\a_S M = \frac{U}{N} \sum_{\k} \left( -\frac{\d n(\epsilon_{\k})}{\d \epsilon_{\k}} \right) M
=U\chi^{0}(q=0)M,
\label{eq:stoner3}
\end{eqnarray}
where $\alpha_S$ is the eigenvalue,
which reaches unity at the critical temperature.
$n(\e)$ is the Fermi distribution function.
In the mean-field approximation, 
$\alpha_S= U\chi^{0}(0)$, where $\chi^0(0)$ is the 
irreducible susceptibility at $\q=0$.
$\alpha_S$ is called the spin Stoner factor.
Equation (\ref{eq:stoner3}) is the simplest example 
of the spin-channel DW equation with $I^s_{kk',q}=U$.

Here, we consider the GL free energy for the 
magnetic transition:
%
\begin{eqnarray}
F_{GL} = a M^2 + \frac{1}{2} b M^4 .
\label{eq:GL0}
\end{eqnarray}
Here, $F_{GL}=\Omega_{\rm MF}+\mu N$,
where $\mu$ is the chemical potential.
Based on Eq. (\ref{eq:stoner1}), the
coefficient $a$ is expressed as
\begin{eqnarray}
a &=& \chi^{0}(0) \left( -1+\frac{1}{\alpha_S} \right) .
\label{eq:GL1}
\end{eqnarray}
This equation is 
consistent with the Stoner theory \cite{stoner}.

For the magnetic state at finite wavevector $\q$,
the eigenvalue of Eq. (\ref{eq:stoner3}) is given as
$\a_S(\q)=U\chi^{0}(\q,0)$,
and the $\q$-dependent coefficient of the GL free energy becomes
$\displaystyle a_\q= \chi^{0}(\q,0) \left( -1+\frac{1}{\alpha_S(\q)} \right)$.
Note that 
$\displaystyle a_\q \approx a_{\q={\bm0}}+\frac12 \sum_\mu^{x,y,z}c_\mu q_\mu^2$,
where $c_\mu=\d_\mu^2a_\q/\d q_\mu^2|_{\q={\bm0}}$.
Here,
$\displaystyle \chi^0(\q,0)= \frac1N \sum_\k 
\frac{n(\e_{\k+\q})-n(\e_{\k})}{\e_{\k}-\e_{\k+\q}}$
is the irreducible susceptibility for general $\q$.
The relation $\a_S>\a_S(\q)$ holds for $\q\ne{\bm0}$
in ferromagnetic metals.

Considering the $T$-dependence of $\chi^0 (0)$, the coefficient 
$a_{\q=\bm{0}}$ at $T=0$ is given as
\begin{eqnarray}
a_{\q=\bm{0}}(T=0)\simeq \frac{\pi^2}{3} B T_{\rm c}^2 ,
\label{eq:GL1-3}
\end{eqnarray}
where $B\equiv D''(0)+{(D'(0))^2}/{D(0)}$,
and $D(0)$ is the density of states at the Fermi energy.
Thus, the necessary condition for the FM transition
($T_{\rm c}>0$) is given as $B<0$
in the mean-field approximation.

However, recent experiments 
have revealed that the ferro-DW order appears even in the case of
$B>0$ in several strongly correlated electron systems 
such as Fe-based superconductors.
Thus, to understand the ferro-DW transition, we have to go beyond the
mean field theory.
In addition, exotic nonlocal DW orders
($\delta t_{ij}^\s$ with $i\neq j$) summarized in Fig. \ref{fig:model}
cannot be explained within the mean-field approximation.
To solve these difficulties, in this paper,
we study the DW phase transition on the bases of the LW theory.


\section{Derivation of DW-equation from Luttinger-Ward theory}
\label{sec:DW}
The Luttinger-Ward (LW) theory in Ref. \cite{LW} provides an exact expression of the 
grand potential $\Omega$, which is applicable for strongly correlated metals unless the
perturbation treatment is violated.
In the first half part of this section,
we discuss the order parameter at $\q={\bm 0}$.
Hereafter, we omit the orbital indices of the Green functions 
and the Coulomb interactions to simplify the expressions, 
because it is straightforward to denote them explicitly.
In the LW theory, the grand potential is expressed as
%
\begin{eqnarray}
\Omega_{\rm{LW}}[G]&=& \Omega_{\rm{F}}[G]+\Phi[G],
\\
\Omega_{\rm{F}}[G]&=&\frac{T}{N} \sum_{k\sigma} \left\{ \ln (G_{k\sigma}) \right.
\nonumber \\
& &\left. -G_{k\sigma}\left( (G^{\rm{\rm{free}}}_{k\sigma})^{-1}-G^{-1}_{k\sigma}  \right) \right\},
\label{eq:omega0}
\end{eqnarray}
where $k\equiv (\k,\e_n)$:
$\k=(k_x,k_y)$ is the wavevector and 
$\e_n=(2n+1)\pi T$ is the fermion Matsubara frequency.
Here, $T\sum_k\cdots \equiv T\sum_{\e_n}\sum_\k\cdots$, and
$G_{k\sigma}$ is the Green function with the self-energy:
$G_{k\sigma}=\left( \left( G_{k\sigma}^{\rm{free}} \right)^{-1} -\Sigma_{k\sigma} \right)^{-1}$.
Now, the self-energy can be divided into 
\begin{eqnarray}
\Sigma=\Sigma^0+\delta t ,
\label{eqn:self-total}
\end{eqnarray}
where $\Sigma^0$ is the ``normal state self-energy''
without any symmetry breaking,
and $\delta t$ is equal to the DW order parameter 
introduced in Sec.\ref{sec:Intro}.
Here, $\Sigma^0$ belongs to $A_{1g}$ symmetry, while 
$\dd \Sigma$ belongs to non-$A_{1g}$ symmetry.
Thus, $\delta t=0$ for $T>T_{\rm c}$.

In Eq. (\ref{eq:omega0}),
$\Phi [G]$ is the Luttinger-Ward function which is
given by calculating the all closed linked skeleton diagrams.
Based on Eq. (\ref{eq:omega0}), we can define $\Omega$ 
as a functional of $\Sigma$ \cite{Potthoff}. 
\begin{eqnarray}
\Omega[\Sigma]=- \frac{T}{N} \sum_{k\sigma}  \ln (-(G_{k\sigma}^{\rm{\rm{free}}})^{-1}+\Sigma_{k\sigma})+P[\Sigma], 
\label{eq:omega}
\end{eqnarray}
where 
$P[\Sigma]$ is considered as the Legendre-transformation of $\Phi[G]$ introduced by Potthoff
\cite{Potthoff}.
\begin{eqnarray}
P[\Sigma]&\equiv & -\frac{T}{N}\sum_{k\sigma} G_{k\sigma} \Sigma_{k\sigma}+\Phi[G].
\end{eqnarray}
In deriving the GL free energy, we have to analyze $\Omega[\Sigma]$ in 
Eq. (\ref{eq:omega}).
Using the Luttinger-Ward function $\Phi$ and Potthoff function $F$
\cite{Potthoff}, 
the self-energy and Green function are respectively expressed as
\begin{eqnarray}
\frac{\dd \Phi[G]}{\dd G_{k\sigma}}
&=& \Sigma_{k\sigma}[G] ,
\\
\frac{\dd P[\Sigma]}{\dd \Sigma_{k\sigma}}
&=& - G_{k\sigma}[\Sigma] .
\end{eqnarray}
Then, the functional derivations of $\Omega[G]$ and $\Omega[\Sigma]$ are respectively given by
\begin{eqnarray}
\frac{\dd \Omega[G]}{\dd G_{k\sigma}}
&=&G_{k\sigma}^{-1} -(G^{\rm{free}}_{k\sigma})^{-1}+\Sigma_{k\sigma}[G] ,
\label{dyson1} \\
\frac{\dd \Omega[\Sigma]}{\dd \Sigma_{k\sigma}}
&=& \frac{1}{(G_{k\sigma}^{\rm{free}})^{-1}-\Sigma_{k\sigma}}-G_{k\sigma}[\Sigma] .
\label{dyson2}
\end{eqnarray}
When $\Omega$ is stationary, the following Dyson equation is satisfied,
\begin{eqnarray} 
\Sigma_{k\sigma}[G]=(G_{k\sigma}^{\rm{free}})^{-1}-G_{k\sigma}^{-1}
\,\,\,\,\,\,\,\,(\mbox{from Eq.(\ref{dyson1})}=0) ,
\label{dysonn1}\\
G_{k\sigma}[\Sigma]=\frac{1}{(G_{k\sigma}^{\rm{free}})^{-1}-\Sigma_{k\sigma}}
\,\,\,\,\,\,\,\,(\mbox{from Eq.(\ref{dyson2})}=0) .
 \label{dysonn2}
\end{eqnarray}
Based on the Luttinger-Ward theory, the ferro $(q=0)$ DW transitions are naturally described
by the self-consistent equation (We call this the DW equation).

\begin{figure}[htb]
\includegraphics[width=.9\linewidth]{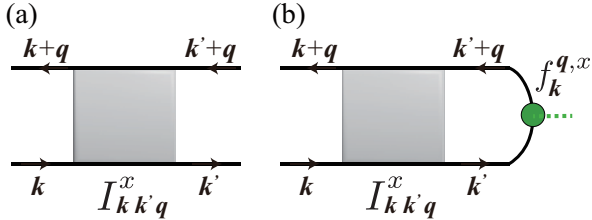}
\caption{
(a) Definition of the irreducible 4-point vertex function $I^x_{kkq}$ ($x=s,c$).
(b) Right-hand side of the linearized DW equation composed of $I^x_{kkq}$.
$f_k^{q,x}$ is the form factor at wavevector $q$.
}
\label{fig:VC-diagram}
\end{figure}

Here, we introduce the  irreducible 4-point vertex $I^{\sigma\sigma'}_{kk'}$
shown in Fig.\ref{fig:VC-diagram} (a).
It is a Jacobian connecting $\Sigma$ and $G$ as
\cite{Potthoff}
\begin{eqnarray}
\frac{\dd \Sigma_{k\sigma}[G]}{\dd G_{k'\sigma'}}&=&I^{\sigma \sigma'}_{kk'} ,
 \label{eq:gamma1}\\
\frac{\dd G_{k\sigma}[\Sigma]}{\dd \Sigma_{k'\sigma'}}&=&\left\{ I^{\sigma \sigma'}_{kk'}
\right\}^{-1} .
 \label{eq:gamma2}
\end{eqnarray}
%
Therefore, the following analytical relation \cite{Potthoff} is obtained:
\begin{eqnarray}
\frac{T}{N}\sum_{k''\sigma''}\frac{\dd \Sigma_{k\sigma}[G]}{\dd G_{k''\sigma''}}\,\,
\frac{\dd G_{k''\sigma''}[\Sigma]}{\dd \Sigma_{k'\sigma'}}=\delta_{kk'}\delta_{\sigma \sigma'} ,
\label{eq:gamma3}
\end{eqnarray}
which is exactly satisfied when $G$ is uniquely determined from $\Sigma$ via one-to-one correspondence.

\begin{figure}[htb]
\includegraphics[width=.9\linewidth]{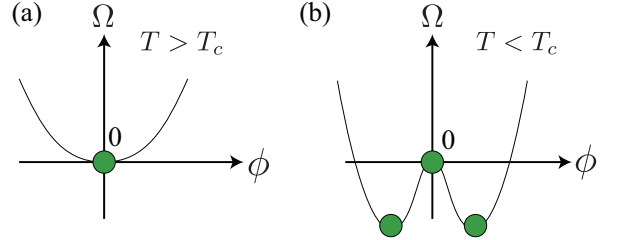}
\caption{
Schematic picture of the GL free energy 
(a) above $T_{\rm c}$ ($a(T)>0$) and (b) below $T_{\rm c}$ ($a(T)<0$).
Stationary points are shown by blue circles.
}
\label{fig:GLpotential}
\end{figure}

The DW equation is derived from the following stationary conditions
\begin{eqnarray}
\left. \frac{\dd \Omega_{\rm{LW}} [G]}{\dd \Sigma_{k\sigma}} \right|_{\Sigma^0}&=&0,
 \,\,\,\,\,\, (\mbox{at any $T$})
\label{dysoonG0}
\\
\left. \frac{\dd \Omega_{\rm{LW}} [G]}{\dd \Sigma_{k\sigma}} \right|_{\bar{\Sigma}}&=&0
 \,\,\,\,\,\,\,\,\, (T<T_{\rm c}),
\label{dysoonG}
\end{eqnarray}
where $\Sigma^0$ is the self-energy without any symmetry breaking, and $\bar{\Sigma}$
is the stationary self-energy after the symmetry breaking.
For $T>T_{\rm c}$, the thermodynamic state is given by Eq. (\ref{dysoonG0}),
which corresponds to the minimum of the free energy 
shown in Fig. \ref{fig:GLpotential} (a).
For $T<T_{\rm c}$, Eq. (\ref{dysoonG}) gives the symmetry breaking state
shown in Fig. \ref{fig:GLpotential} (b).
(Eq. (\ref{dysoonG}) corresponds to the unstable extremum.)

By using Eq.(\ref{dysoonG}) and (\ref{dyson2}),
\begin{eqnarray}
\left. \frac{\dd \Omega_{\rm{LW}} [G]}{\dd \Sigma_{k\sigma}}\right|_{\bar{\Sigma}}
&=&
-G_{k\sigma}^2\Sigma_{k\sigma}+\frac{\dd \Phi[G]}{\dd \Sigma_{k\sigma}} 
\nonumber \\
&=&-G_{k\sigma}^2\Sigma_{k\sigma}
+\frac{T}{N}\sum_{k'\sigma'}\frac{\dd G_{k'\sigma'}}{\dd \Sigma_{k\sigma}} 
\frac{\dd \Phi[G]}{\dd G_{k'\sigma'}},
\end{eqnarray}
where $\frac{\dd G_{k'\sigma'}}{\dd \Sigma_{k\sigma}}=G^2_{k\sigma} \delta_{kk'}$.
Thus, the stationary condition of Eq. (\ref{dysoonG}) is rewritten as
\begin{eqnarray}
\left. \frac{\dd \Phi[G]}{\dd G_{k\sigma}}\right|_{G^0}&=&\Sigma_{k\sigma}^{0} ,
\label{eq:sta1}\\
\left. \frac{\dd \Phi[G]}{\dd G_{k\sigma}}\right|_{\bar{G}}
&=&\bar{\Sigma}_{k\sigma} ,
\label{eq:sta2}
\end{eqnarray}
where ${\hat G}^0=(\{{\hat G}^{\rm free}\}^{-1}-{\hat \Sigma}^0)^{-1}$.
Equations (\ref{eq:sta1}) and (\ref{eq:sta2}) compose the 
``exact DW equation'' that describes the DW state below $T_{\rm c}$.
Then, the order parameter is 
$\delta t_{k\sigma}= \bar{\Sigma}_{k\sigma}-\Sigma_{k\sigma}^{0}$.

It is noteworthy that superconducting gap equation is 
derived from the stationary condition of the LW function 
in the SC state $\delta \Phi[G,F]$, 
where $F$ ($G$) is the anomalous (normal) Green function
 \cite{Yanase}.
Thus, the derived DW equation for the form factor, 
Eqs. (\ref{eq:sta1}) and (\ref{eq:sta2}),
is well-founded comparable with the well-known superconducting-gap equation.

Next, we derive the linearized DW equation with respect to $\delta t$.
By subtracting Eq. (\ref{eq:sta1}) from (\ref{eq:sta2}) , we obtain
\begin{eqnarray}
\dd t_{k\sigma}&=& \frac{T}{N}\sum_{k'\sigma'} \left. \frac{\dd^2 \Phi[G]}{\dd G_{k'\sigma'} \dd G_{k\sigma}} \right|_{\bar{G}} \dd G_{k'\sigma'} \nonumber \\
&=& \frac{T}{N}\sum_{k'\sigma'} \left. \frac{\dd^2 \Phi[G]}{\dd G_{k'\sigma'} \dd G_{k\sigma}} \right|_{G^0}\dd G_{k'\sigma'} +O(\dd G^2) ,
\label{eqn:DW-expand}
\end{eqnarray}
where $\dd G\equiv \bar{G}-G^0$.
Here, we rewrite $\dd t_{k\sigma}$ as
\begin{eqnarray}
\dd t_{k\sigma} \equiv  \phi \cdot f_{k\sigma} ,
\end{eqnarray}
where $\phi$ is a real parameter, and $f_{k}^{q}$ is 
the normalized order parameter that belongs to one of the 
irreducible representations in non-$A_{1g}$ symmetry.
It is convenient to set $\max_k |f_{k\sigma}|=1$ 
because the relation $\phi=\max_k|\dd t_{k\sigma}|$ holds.
Thus, we derive the following ``linearize DW equation for $\q={\bm0}$''
by introducing the eigenvalue $\lambda$ to the 
left-hand side of Eq. (\ref{eqn:DW-expand});
\begin{eqnarray}
\lambda f_{k\sigma}= \frac{T}{N} \sum_{k'\sigma' }  I^{\sigma \sigma'}_{kk'}
(G^{0}_{k'\sigma'})^2  f_{k'\sigma'} ,
\label{eq:cons06q0}
\end{eqnarray} 
where we denote
$I^{\sigma \sigma'}_{kk'} \equiv \left. I^{\sigma \sigma'}_{kk'} \right|_{\Sigma^0}$
to simplify the notation.
In Eq. (\ref{eq:cons06q0}),
the largest eigenvalue $\lambda$ reaches unity at $T=T_{\rm c}$,
and its eigenvector gives the form factor of the DW state.

The linearized DW equation can be generalized for finite $q$ orders as follows.
First, we consider the DW order with the wavevector $\bm{q}=\bm{g}\cdot m/n$ where $\bm{g}$ is the reciprocal lattice vector, and $m$, $n$ 
are integers ($0\le m< n$).
Then, we can introduce the $n\times n$-matrix Green function
$G^{lm}_k=\langle k+lq| \hat{G} |k+mq \rangle$ where $l,m=0 \sim n-1$.
In this case, Eq. (\ref{eq:sta2}) becomes
\begin{eqnarray}
\left. \frac{\dd \Phi[G]}{\dd G_{k\sigma}}\right|_{\bar{G}^{lm}}&=&\bar{\Sigma}_{k\sigma}^{lm} .
\end{eqnarray}
Hereafter, we drop the overlines of $\bar{G}$ and $\bar{\Sigma}$
to simplify the notation.

Here, we adopt the extended Brillouin zone scheme for $\vec{k}$ to simplify the explanation.
After that, the DW equation can be linearized with respect to the 
$\q$-linear term in $\dd t$ given by $\dd t^{m+1,m}\equiv \dd t^{q}_k$.
By introducing the $q$-dependent eigenvalue, we obtain the following 
``linearize DW equation for general $\q$'',
\begin{eqnarray}
\lambda_q   f_{k\sigma}^q= \frac{T}{N} \sum_{k'\sigma' }  I^{\sigma \sigma'}_{kk'q}
G^{0}_{k'\sigma'} G^{0}_{k'+q\sigma'} f_{k'\sigma'}^q ,
\label{eq:cons06}
\end{eqnarray}
where $q\equiv (\q,\w_l)$:
$\q=(q_x,q_y)$ is the wavevector and $\w_l=2l\pi T$
is the boson Matsubara frequency.
The condition $\lambda_q=1$ brings the DW transition temperature $T_{\rm c}$
with wavevector $q$, which can be interpreted as the particle-hole (p-h) gap equation.
The form factor of the eigenvalue equation (\ref{eq:cons06}) 
contains the uncertainty of the phase factor $e^{i\theta}$.
The correct phase $\theta$ is uniquely fixed by following the 
Hermitian condition in Eq. (\ref{eq:Hermite}).


The DW equation (\ref{eq:cons06}) is further simplified 
by introducing the spin (s) and charge (c) channel functions
in the absence of the spin-orbit interaction \cite{Kontani-sLC}:
\begin{eqnarray}
f^{c(s)}&=&f_{\uparrow}+(-)f_{\downarrow}.
%
%
\\
I^{c(s)} &=&I^{\uparrow \uparrow}+(-)I^{\uparrow \downarrow} .
\end{eqnarray}

Finally, we derive the simplified
``linearized DW equation for $x \ (=s,c)$ channel form factor at $q$''
from Eq. (\ref{eq:cons06}) as follows:
\begin{eqnarray}
\lambda^{x}_q \,\, f_{k}^{q,x}= \frac{T}{N} \sum_{k' }  I^{x}_{kk'q}
G^0_{k'} G^0_{k'+q} \,\, f_{k'}^{q,x} .
\label{eq:cons06cs}
\end{eqnarray}
The right-hand side of Eq. (\ref{eq:cons06cs})
is shown in Fig.\ref{fig:VC-diagram} (b).
Therefore, we derived the exact expression of the DW equation
in Eq. (\ref{eq:cons06}) or Eq. (\ref{eq:cons06cs})
composed of the true irreducible four-point vertex $I$ and the self-energy.

We can derive the expression of $\lambda^{x}_q$ from 
Eq. (\ref{eq:cons06cs}) as
\begin{eqnarray}
\lambda^{x}_q = X^x_q/\chi^{0f}(q) ,
\label{eq:lambda_q}
\end{eqnarray}
where $X$ is given as
\begin{eqnarray}
X^x_q=\frac{T^2}{N^2} \sum_{k,k'} (f_{k}^{q,x})^* G^0_{k} G^0_{k+q}
I^{x}_{kk'q} G^0_{k'} G^0_{k'+q} f_{k'}^{q,x} ,
\label{eq:X_q}
\end{eqnarray}
and $\chi^{0f}(q)$ is the irreducible susceptibility 
with the form factor $f_{k\sigma}^q$:
\begin{eqnarray}
\chi^{0f} (q) &\equiv & -\frac{T}{N} \sum_{k \sigma} 
f_{k\sigma}^q G^{0}_{k+q \sigma}G^{0}_{k \sigma} f_{k\sigma}^{-q}.
\label{eq:chi0f}
\end{eqnarray}
From Eq. (\ref{eq:lambda_q}),
the relation $\lambda^{x}_q \propto X^x_q$ is obtained 
because $\chi^{0f}(q)$ is nearly $T$-independent.

It is noteworthy that the DW equation introduced by Onari and Kontani {\it et al.} in Refs. \cite{Onari-FeSe,Kawaguchi-CDW,Tsuchiizu-CDW,Tazai-JPSJ-fRG}, which has been applied to iron-based and cuprate superconductors, is derived from the exact DW equation given in Eq. (\ref{eq:cons06}). The detailed derivation is shown in Appendix A.

Before closing this section, we reproduce the Stoner theory by 
applying the mean-field approximation to Eq. (\ref{eq:cons06cs}).
In the mean-field theory,
\begin{eqnarray}
I^{s}_{kk'q}=-U , \,\,\,\,\,\,\,\,\,\,\,\,  \dd \Sigma^{q,s}_{k}=M\delta_{q,0} .
\end{eqnarray}
Therefore, the  Eq. (\ref{eq:cons06cs}) is given by
\begin{eqnarray}
\lambda^{s} M = -\frac{T}{N} \sum_{k'} G^0_{k'}G^0_{k'} U M
=U \chi^0  M .
\end{eqnarray}
Thus, the eigenvalue for the ferromagnetic transition corresponds to  
the Stoner factor $\alpha_{S}$:
\begin{eqnarray}
\lambda^s=U \chi^0 \equiv\alpha_{S} .
\end{eqnarray}
%

\section{Derivation of GL free energy based on the linearized DW equation}
\label{sec:LW}

\subsection{GL free energy for DW state with form factor $f$}
Here, we derive the expression of the Ginzburg-Landau (GL) \rm{free} energy
based on the DW-equation. The GL \rm{free} energy due to the ferro-DW transition
($\dd t_{k\sigma}=\phi f_{k\sigma}$) is expressed as
\begin{eqnarray}
\Omega_{\rm{DW}}(T,\mu,\phi)= a(T) \phi^2 +  \frac12 b \phi^4 .
\label{eq:omega_tenkai2}
\end{eqnarray}
Its schematic picture above $T_{\rm c}$ ($a(T)>0$) and 
that below $T_{\rm c}$ ($a(T)<0$) are shown in 
Figs. \ref{fig:GLpotential} (a) and (b), respectively.

Note that the coefficients $a$ and $b$ are functional of the form factor $f_{k\sigma}$.
By using the $\Omega[\Sigma]$ defined in Eq.(\ref{eq:omega}),
the coefficient $a$ is calculated from the 
second functional derivation of $\Omega[\Sigma]$ as
\begin{eqnarray}
&&\Omega[\Sigma^0+\dd t]-\Omega[\Sigma^0] 
\nonumber \\
&& \ \ \ \ \ \approx
\frac{T}{N}\sum_{k\sigma k'\sigma'}  \left. \frac{\dd^2 \Omega[\Sigma]}{\dd \Sigma_{k'\sigma'}\dd \Sigma_{k\sigma}}\right|_{\Sigma=\Sigma^0} \dd t_{k'\sigma'} \dd t_{k\sigma} . 
\label{eq:omega_tenkai3}
\end{eqnarray}
By using Eq. (\ref{dyson2}), we obtain 
\begin{eqnarray}
\frac{\dd^2 \Omega[\Sigma]}{\dd \Sigma_{k'\sigma'}\dd \Sigma_{k\sigma} }&=&
\frac{\dd}{\dd \Sigma_{k'\sigma'}}\left\{\frac{1}{(G_{k\sigma}^{\rm{free}})^{-1}-\Sigma
_{k\sigma}}-G_{k\sigma}[\Sigma]\right\} \nonumber \\
&=& G_{k\sigma}^2 \delta_{kk'}-\frac{\dd G_{k\sigma}[\Sigma]}{\dd \Sigma_{k'\sigma'}} 
\nonumber \\
&=& (G^0_{k\sigma})^2\delta_{kk'}- \left\{I_{kk'}^{\sigma \sigma'} \right\}^{-1} ,
\label{eq:omega_tenkai3-2}
\end{eqnarray}
where we used the relation in Eq. (\ref{eq:gamma2}).
%
Therefore, the Eq. (\ref{eq:omega_tenkai3}) is rewritten as
\begin{eqnarray}
&&\Omega[\Sigma^0+\dd t]-\Omega[\Sigma^0] 
\approx
\frac{T}{N}\sum_{k\sigma} (G^0_{k\sigma})^2 (\dd t_{k\sigma})^2  
\nonumber \\
&& \ \ \ \ \ \ 
- \frac{T}{N}\sum_{k\sigma k'\sigma'}  \left\{I_{kk'}^{\sigma \sigma'} \right\}^{-1}  \dd t_{k\sigma} \dd t_{k'\sigma'} .
\label{eq:omega_tenkai4}
\end{eqnarray}
Here, we recall that the order parameter 
$\dd t_{k\sigma}=\phi\cdot f_{k\s}$
is determined  by using the DW equation (\ref{eq:cons06}).
%
By using Eq. (\ref{eq:cons06}) together with 
Eqs. (\ref{eq:gamma1})-(\ref{eq:gamma3}),
Eq. (\ref{eq:omega_tenkai4}) is rewritten as
\begin{eqnarray}
\Omega[\Sigma^0+\dd t]-\Omega[\Sigma^0] &\simeq & 
\left( 1-\frac{1}{\lambda}\right)
\frac{T}{N}\sum_{k\sigma} (G^0_{k\sigma})^2 f_{k\sigma}^2 \phi^2 \nonumber \\
&=&-2\chi^{0f}(0) \left\{ 1-\frac{1}{\lambda} \right\} \phi^2 ,
\label{eq:delsigma2}
\end{eqnarray}
where the factor $2$ originates from the spin degeneracy.

We can derive the GL free energy for the order parameter at
nonzero wavevector $q$ by considering the large unit cell 
as we discussed in Sect. \ref{sec:DW}.
Thus, the coefficient $a$ defined in the Gibbs \rm{free} energy in
Eq. (\ref{eq:omega_tenkai2}) is obtained as
\begin{eqnarray} 
a_q(T)&=& -2\chi^{0f}(q) \left(1-\frac{1}{\lambda_q}\right).
\label{eq:delsigma2-2}
\end{eqnarray}
As a result, we obtain the exact expression for coefficient $a_q$
by using the eigenvalue $\lambda_q$ in the DW equation.
The obtained general expression in  
Eqs. (\ref{eq:chi0f}) and (\ref{eq:delsigma2-2}) are 
meaningful to discuss the DW transition.

Finally, we stress that the Potthoff's Legendre-transformation
of the LW formalism \cite{Potthoff}
is necessary to derive the correct GL free energy expression.
In Appendix B, we explain that the expansion of (\ref{eq:omega0})
with respect to $\delta t$ leads to an inaccurate expression.


\subsection{GL free energy for BCS Superconductivity}
\label{sec:BCS case}
Here, we consider the GL equation for the spin-singlet superconductivity. 
Here, we express the spin-singlet SC gap function as
$\Delta_k = \psi\cdot f_k$,
where $f_\k$ is the normalized form factor.
Based on the LW theory for the superconducting states, 
we can derive that the second order GL parameter is given by
\begin{eqnarray}
a (T)&=& -2\chi_{\rm pp}^{0\psi} (0) \left(1-\frac{1}{\lambda_{\rm sc}}\right), 
\label{eq:delsigma2BCS} \\
\chi_{\rm pp}^{0\psi} (0) &= & \frac{T}{N} \sum_{k} | \psi_k|^2 
G^{0}_{k} G^{0}_{-k} ,
\label{eq:chi0fBCS} 
\end{eqnarray}
where $\lambda$ is the eigenvalue of the linearized gap equation given by
\begin{eqnarray}
\lambda_{\rm sc} f_k= \frac{T}{N} \sum_{k'} V_{kk'} G^{0}_{k'} G^{0}_{-k'} f_{k'} ,
\label{eq:chi0fBCS3} 
\end{eqnarray}
where $V_{kk'}=\left. \frac{\dd^2 \Phi}{\dd F \dd F^{\dagger}}\right|_{\Delta=0}$. Here, $F$ and $F^{\dagger}$ are anomalous Green functions.
The derivation of Eq. (\ref{eq:delsigma2BCS}) is essentially the same as that for the DW transition given in previous sections. We can show that the relationship (\ref{eq:delsigma2BCS}) is also valid for the spin-triplet superconductivity.


\section{numerical analysis of nematic state in FeSe}

In this section,
we explain the important unsolved problems in FeSe,
which is one of the most famous Fe-based superconductors.
We try to understand the following key topic on the nematicity,
for both above and below the nematic transition temperature $T_{\rm c}$, 
based on a unified theory.
{\bf (i)} Lifshitz transition below $T_{\rm c}$.
{\bf (ii)} Nematic susceptibility above and below $T_{\rm c}$.
{\bf (iii)} Specific heat jump at $T=T_{\rm c}$.
In FeSe, $T_{\rm c}$ corresponds to the 
structure transition temperature $T_S$.

In previous sections,
we derived the exact expressions of the 
linearized DW equation in Eq. (\ref{eq:cons06cs}) for $T>T_{\rm c}$
and the full DW equation in 
Eqs. (\ref{eq:sta1})-(\ref{eq:sta2}) for $T<T_{\rm c}$.
Here, we solve these equations for FeSe
based on the one-loop approximation for the LW function, $\Phi_{\rm FLEX}$,
derived in Appendix A.
We include the normal state (=without order parameter)
self-energy $\Sigma^0$ into the DW equations
because it is necessary to satisfy the stationary condition
 (\ref{eq:sta1})-(\ref{eq:sta2}),
although it was dropped in previous studies
\cite{Onari-FeSe}.

We study a realistic $d+p$ orbital Hubbard model
with on-site multiorbital Coulomb interaction $U$ for FeSe:
\begin{eqnarray}
H = H_0 + r H_U ,
\label{eqn:Ham}
\end{eqnarray}
%
where $H_0$ is the $d$+$p$ orbital tight-binding model for FeSe, 
and $H_U$ is the $d$-orbital Coulomb interaction for $d$+$p$ orbital model 
given by the constrained-random-phase-approximation (cRPA) method. 
The matrix elements in $H_U$ are composed of the 
intra-orbital Coulomb repulsion $U_{l,l}$, 
the inter-orbital Coulomb repulsion $U_{l,m}$ $(l\ne m)$, 
and the exchange term $J_{l,m}$, as we explain in Appendix C.
\color{black}


In Eq. (\ref{eqn:Ham}), 
$r (< 1)$ is the reduction factor of $H_U$ 
that represents the screening due to $p$ orbitals. 
According to Ref. \cite{Miyake}, 
the averaged intra-orbital $U_{\rm av}\sim7eV$ is reduced to $\sim4$eV 
due to the screening effect by $p$-orbitals. 
In the present study, we set $r=0.3\sim0.4$, where $T_c$ increases with $r$. 
In contrast, $T_c$ slowly decreases with $r$ for $r>0.5$. 
Thus, obtained $T_c$ depends on $r$, while the symmetry and the form factor 
of the nematic order is insensitive to the choice of $r$. 
We note that the relation $\alpha_S<1$ is satisfied for any $r$ 
in the present two-dimensional FeSe model because the FLEX approximation 
satisfies the Mermin-Wagner theorem 
\cite{Mermin}. 
This fact is favorable for realizing the nonmagnetic nematic state ($\lambda>1$ and $\alpha_S<1$).
\color{black}

In the present numerical study for FeSe, 
we use $64\times64$ $\k$-meshes, 
and 4096 or 8192 Matsubara frequencies.
Figure \ref{fig:heat-FeSe} (a)
represents the Fermi surface (FS) of FeSe model.
We derive the normal self-energy $\Sigma^0$
by applying the FLEX approximation.
In the case of $r=0.36$, 
the obtained orbital-dependent mass-enhancement factors at $T=10$meV
are about $z_{xy}^{-1}\sim 5$ and $z_{xz,yz}^{-1}\sim3$, respectively.
The Stoner factor is about $0.9$ and its $T$-dependence is very weak.

\subsection{Above $T_{\rm c}$: linearized DW equation analysis}

First, we analyze the multiorbital Hubbard model for FeSe
based on the linearized DW equation in Eq. (\ref{eq:cons06cs}),
with the kernel function in Eq. (\ref{eqn:IcS}).
Here, we incorporate the normal state self-energy $\Sigma^0$
given in Eq. (\ref{eqn:self0}) into the DW equation 
to perform the conserving approximation.
$\Sigma^0$ is significant to derive realistic $T_{\rm c}$
and beautiful CW/non-CW behaviors of $\chi_{\rm nem}$,
although it has been dropped in our previous analyses.


Here, we discuss the kernel function in Eq. (\ref{eqn:IcS}).
The first line in Eq. (\ref{eqn:IcS}) gives the Hartree term, 
Maki-Thompson (MT) term, and the second and the third lines in 
Eq. (\ref{eqn:IcS}) give the Aslamazov-Larkin (AL) terms.
Both MT and AL terms cause important
 ``fluctuation-induced interaction for the DW''.
In Fe-based SCs, the nematic order mainly originates from the AL terms,
which represent the ``interference between paramagnons''
\cite{Onari-SCVC,Onari-FeSe,Yamakawa-PRX2016,Chubukov-AL}.
The MT term is also important to induce the characteristic 
sign-reversing in the form factor in $\k$-space
\cite{Onari-FeSe,Chubukov-nematic-rev}.
On the other hand, the cLC orders in 
geometrically frustrated Hubbard models
mainly originate from the the MT terms 
\cite{Tazai-cLC}.
Note that the MT terms induce striking
non-Fermi-liquid transport phenomena near the QCPs
\cite{Kontani-ROP}.

\begin{figure}[htb]
\includegraphics[width=.98\linewidth]{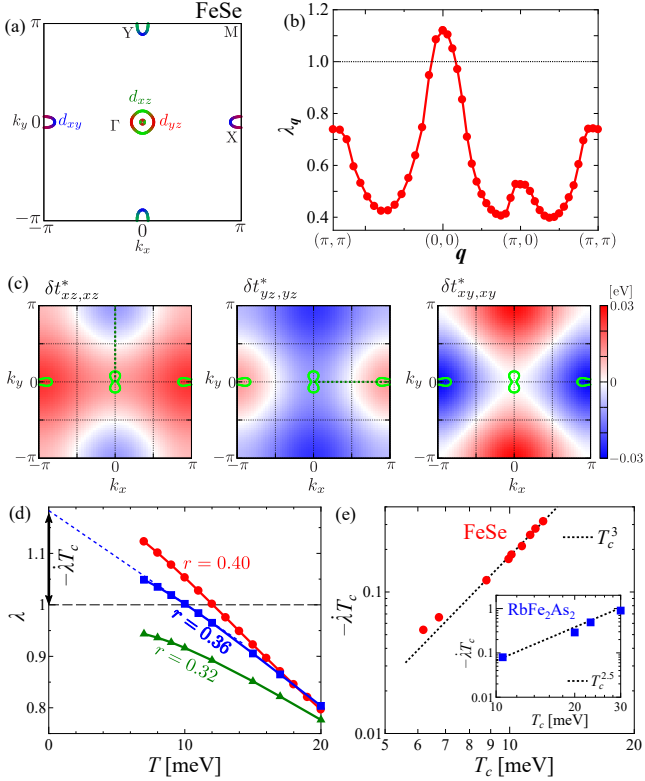}
\caption{
(a) FS of FeSe model.
Here, the weights of the $xz$, $yz$, and $xy$ orbitals 
are shown by green, red, and blue colors.
(b) Obtained eigenvalue $\lambda_\q$ at $T=5$meV and $r=0.40$
derived from the linearized DW equation.
The fact that $\lambda_\q$ exhibits the maximum at $\q={\bm0}$
means the emergence of the ferro-DW order.
(c) Renormalized form factor at $\q={\rm0}$, 
$\delta t_{\k,m,m}^*$ with $m=m'=xz$, $yz$ and $xy$,
derived from the full DW equation at $T=5$meV;
see Fig. \ref{fig:heat-FeSe2}.
The deformed FS in the nematic state is shown.
(d) $T$-dependence of $\lambda_{\q={\bm0}}(T)$ for $r=0.32$, $0.36$ and $0.40$.
(e) Obtained $R=(-\dot{\lambda}T_{\rm c})$ as a function of $T_{\rm c}$ for $r=0.40\sim0.32$.
The relation $R\propto T_{\rm c}^b$ with $b\approx3$ holds.
The inset shows $R$ in the RbFe$_2$As$_2$ model.
Note that $R\approx1$ ($b=0$) in the BCS superconductivity.
}
\label{fig:heat-FeSe}
\end{figure}

Figure \ref{fig:heat-FeSe} (b) shows the 
$\q$-dependence of the largest charge-channel 
eigenvalue $\lambda_\q^c$ at $r=0.40$ and $T=5$meV 
derived from the DW equation.
(Below, we drop the superscript $c$ of $\lambda_\q^c$ for simplicity.)
The obtained $\lambda_\q$ exhibit the maximum at $\q={\bm0}$
because the convolution of two $\chi_q^s$'s,
$C_q\equiv \sum_p\chi^s_p\chi^s_{p+q}$,
included in the AL-type VCs include is largest at $q=0$.
Here, $\lambda_{\q={\bm0}}$ exceeds unity, 
and typical transition temperature 
$T_{\rm c} \ (\sim100{\rm K})$ in Fe-based SCs
is reproduced by including $\Sigma^0$.
We note that similar results were obtained in Fig. S4 (a)
in Ref. \cite{Onari-smectic} by taking $\Sigma^0$ into account.

The obtained form factors belong to $B_{1g}$ symmetry,
which is shown in Fig.\ref{fig:heat-FeSe} (c):
The $xz,yz$-orbital form factors express the 
$\k$-dependent orbital polarization,
which has been reported by previous DW equation studies without $\Sigma^0$
\cite{Onari-FeSe}.
The obtained $xz,yz$-orbital polarization elongates the hole-pocket
along the $k_y$-axis as experimentally reported in Refs.
\cite{FeSe-ARPES-Suzuki,FeSe-ARPES-Zhang}.
In addition, the $xy$-orbital form factor the represents the $d$-orbital bond order 
$f \propto \cos k_x- \cos k_y$ emerges at the same $T_{\rm c}$.
This $d$-wave order leads to the disappearance of an electron-pocket around Y-point 
\cite{FeSe-Lif1,FeSe-Lif2,Eremin}.
Thus, experimentally-observed ferro-nematic order in FeSe is naturally obtained.
We stress that the coexistence of the $xz,yz$-orbital order
and the $d_{x^2-y^2}$-wave bond order on $xy$-orbital
was already reported in Fig. S3 (a)-(d)
in Ref. \cite{Onari-smectic}.
In addition, the $d_{xy}$-wave bond order on $xy$-orbital 
has been studied in RbFe$_2$As$_2$ in Ref. \cite{Onari-B2g}.

We comment that a simple $A_{1g}$ symmetry order that accompanies 
the net charge order is suppressed by the Hartree term 
in the kernel function.
In contrast, the $B_{1g}$ symmetry order in 
Fig. \ref{fig:heat-FeSe} (c) is free from the suppression
by the Hartree term due to sign reversal in the form factor.

Figure \ref{fig:heat-FeSe} (d) shows the  
temperature-dependence of $\lambda_{\q={\bm0}}(T)$ for $r=0.40\sim0.32$.
At higher temperatures ($T\gtrsim10$meV),
$\lambda_{\q={\bm0}}(T)$ exhibits the $T$-linear behavior.
The nematic susceptibility above $T_{\rm c}$ 
is $\chi_{\rm nem}= \chi^{0f}(0)(1-\lambda_{\q={\bm0}})^{-1}$,
as proved theoretically in Ref. \cite{Onari-B2g}.
Therefore, it is confirmed that the 
experimental CW behavior of $\chi_{\rm nem}$ 
at higher temperature is naturally explained in the present theory.
The deviation from the CW behavior at lower temperatures will be discussed 
in Sect. \ref{sec:Non-Curie-Weiss}.

Figure \ref{fig:heat-FeSe} (e) shows the obtained
$R=(-\dot{\lambda}T_{\rm c})$ as a function of $T_{\rm c}$.
Here, $\displaystyle \dot{\lambda} \equiv \left.\frac{d\lambda}{dT}\right|_{T_{\rm c}}$, 
$R$ approximately corresponds to $\lambda_{\q={\bm0}}(T=0)-1$, and
$R$ becomes very small when $T_{\rm c}\ll 10$meV.
The reason is the recovery of the 
Fermi-liquid behavior $\lambda(0)-\lambda(T)\propto T^2$
because the system is far from the magnetic QCP.
Also, the relation $R\propto T_{\rm c}^b$ ($b\sim3$) 
is satisfied in the present study.
In Sect. \ref{sec:Thermodynamic},
we will explain that $R$ is proportional to the jump in the heat capacity at $T_{\rm c}$.


\subsection{Below $T_{\rm c}$: full DW equation analysis}

Next, we analyze the full DW equation given as
Eqs. (\ref{eq:sta1})-(\ref{eq:sta2}) for $T<T_{\rm c}$.
We safely assume the uniform ($\q={\bm0}$) order parameter 
because $\lambda_\q$ takes the largest value at $\q={\bm0}$
as found in Fig. \ref{fig:heat-FeSe} (b).
The aim of this subsection is to explain the essential 
properties of the nematic state ($T<T_{\rm c}$) in FeSe 
based on the paramagnon interference mechanism.

Now, we explain the procedure of the numerical study in detail:
The total self-energy is given in Eq. (\ref{eqn:self-total}),
where $\Sigma^0$ is the normal self-energy without any symmetry breaking
given by Eq. (\ref{eqn:self0}).
Here, we calculate $\Sigma^0$ at each $T$
by subtracting its static and Hermitian part, 
$\Sigma^{0,{\rm H}}(\k)\equiv(\Sigma^0(\k,+i\delta)+\Sigma^0(\k,-i\delta))/2$,
in order to fix the shape of the FS
\cite{Text-SCVC}.
Next, we derive the symmetry breaking part $\delta t$ self-consistently
based on the following procedure: 
(a) We first calculate 
$\displaystyle S_k\equiv \frac{T}{N}\sum_q G_{k+q}[\Sigma] W_q[\Sigma]$,
where $G_k[\Sigma]$ and $W_q[\Sigma]$ are functions of the total self-energy.
(b) Next, we derive $\delta t$ as
\begin{eqnarray}
\delta t_k =(1-P_{0})S_k ,
\label{eqn:full-DWeq-pro}
\end{eqnarray}
where $P_{0}$ is the projection operator for 
the totally-symmetric ($A_{1g}$) channel.
(c) The total self-energy is given as $\Sigma=\Sigma_0+\delta t$.
We repeat (a)-(c) till $\delta t$ converges.

It is easy to show that the $\delta t$-linear term of $S_k$ gives
the right-hand side of the linearized DW equation (\ref{eq:cons06cs})
with the kernel function in Eq. (\ref{eqn:IcS}).
Thus, the full-DW equation is equivalent to the linearized-DW equation 
when $\delta t$ is very small.

\begin{figure*}[htb]
\includegraphics[width=.6\linewidth]{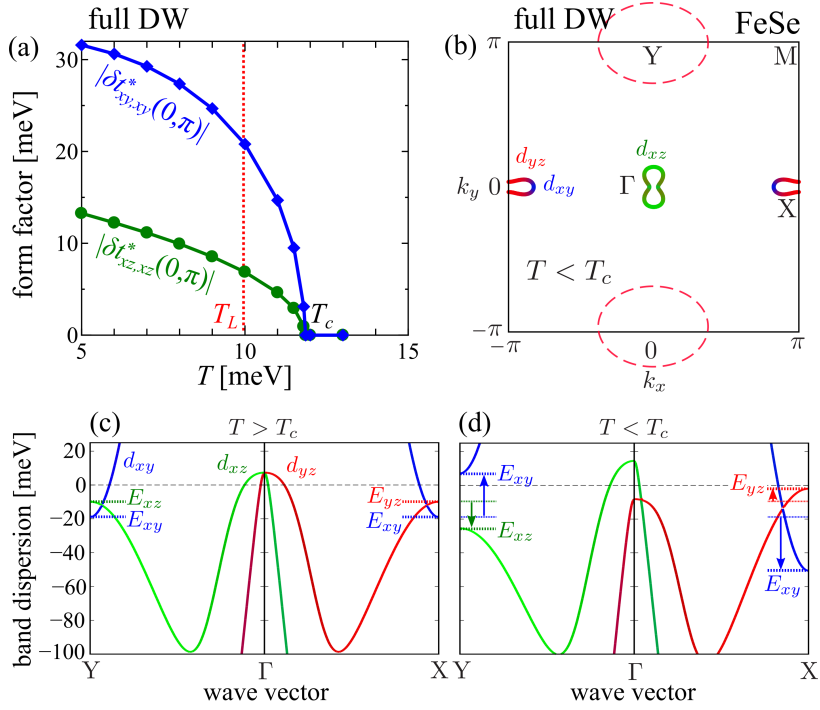}
\caption{
(a) Obtained renormalized symmetry breaking self-energy 
$\delta t_{m,m}^*(\k)$ for $m=xz,xy$ at Y-point,
derived from the full DW equation at $r=0.40$.
The second-order transition occurs at $T_{\rm c}=12$meV,
which is consistent with the linearized DW equation analysis 
in Fig. \ref{fig:heat-FeSe} (d).
$T_L\ (=9{\rm meV})$ is the Lifshitz transition temperature.
(b) FS in the nematic state at $T=5{\rm meV} \ (<T_L)$.
The electron-pocket around Y-point disappears due to the $xy$-orbital form factor.
In addition, the hole-pocket is elongated along the $k_y$-axis
due to the $xz,yz$-orbital polarization with the sign-reversal in $\k$-space.
(c) Renormalized band structure in the normal state.
(d) Renormalized band structure in the nematic state.
}
\label{fig:heat-FeSe2}
\end{figure*}

Figure \ref{fig:heat-FeSe2} (a) represents the 
obtained renormalized order parameter
$\delta t_m^*(\k)=z_{m}\delta t_m(\k)$ 
for $m=xz$ and $m=xy$ at Y-point.
The obtained $T_{\rm c}=12$meV completely coincides with that 
given by the linearized DW equation.
The nematic order occurs as the second-order, and 
the averaged order parameter
$\delta t_{\rm av}^*\equiv (|\delta t_{xz}^*|+|\delta t_{xy}^*|)/2$
at Y-point is about $2T_{\rm c}$ at $T=5$meV.
Thus, the present theory gives the ratio 
$\delta t_{\rm av}^*/T_{\rm c}\sim2$,
which is similar to the ratio $\Delta_0^*/T_{\rm c}^{\rm SC}\sim2$ in the BCS theory.
Thus, both the development of $\chi_{\rm nem}$ above $T_{\rm c}$
and the nematic order parameter below $T_{\rm c}$
are well explained by the present theory.

The relation $2|\delta t_{xz}^*|\approx|\delta t_{xy}^*|$
in Fig. \ref{fig:heat-FeSe2} (a) 
indicates that both the $(xz+yz)$-orbitals and $xy$-orbital
equally contribute to the nematic order.
In Appendix D,
we explain the relative phase between 
$xz$-orbital and $xy$-orbital form factors,
$\delta t_{xz,xz}(0,\pi)\delta t_{xy,xy}(0,\pi)<0$,
on the basis of the Ginzburg-Landau analysis.
This relation is significant for the Lifshitz transition 
below $T_{\rm c}$ as we will explain below.


Due to the nematic order parameter,
the FS with $C_4$-symmetry in Fig. \ref{fig:heat-FeSe} (a)
is deformed to the $C_2$-symmetry FS 
depicted in Fig. \ref{fig:heat-FeSe2} (b) at $r=0.40$ and $T=5$meV.
The corresponding band-dispersions in the normal state 
and that in the nematic state are shown in  
Figs. \ref{fig:heat-FeSe2} (c) and (d), respectively.
They are renormalized by the factor $z\sim5$ for $xy$ orbital
and $z\sim3$ for $xz$,$yz$ orbitals.
The original band-dispersion is shown in 
Fig. \ref{fig:band-original} in Appendix C.
The hole-pocket is elongated along the $k_y$-axis
due to the $xz,yz$-orbital polarization 
\cite{Onari-FeSe,FeSe-ARPES-Suzuki,FeSe-ARPES-Zhang}.
Interestingly,
the electron-pocket around Y-point disappears in the nematic state
due to the $d_{x^2-y^2}$-wave form factor on the $xy$-orbital.
This nematic Lifshitz transition 
has been confirmed by many angle-resolved photoemission spectroscopy 
(ARPES) studies \cite{FeSe-Lif1,FeSe-Lif2}.
The relative phase between two form factors,
$\delta t_{xz,xz}(0,\pi)\delta t_{xy,xy}(0,\pi)<0$,
originates from the kinetic energy gain due to the 
pseudogap formation by the Lifshitz transition.

In addition, in the nematic phase,
the hole-pocket is elongated along the $k_y$-axis
due to the $xz,yz$-orbital polarization with the sign-reversal in $\k$-space.
This has been also confirmed by ARPES studies
\cite{FeSe-ARPES-Suzuki,FeSe-ARPES-Zhang}.
Thus, experimental key findings in the nematic states
are satisfactorily reproduced by the present 
``paramagnon interference mechanism''
\cite{Onari-SCVC,Yamakawa-PRX2016,Onari-FeSe,Tazai-JPSJ-fRG}.

\subsection{Connection between above and below $T_{\rm c}$}

In the previous subsection, we derived the nematic self-energy
$\Sigma\equiv \Sigma^0+\delta t$ based on the full DW equation.
The derived nematic state corresponds to the 
stationary point of the LW grand potential $\Omega_{\rm LW}[\Sigma]$,
as proved in Sect. \ref{sec:DW}.
Because the obtained nematic state is thermodynamically stable,
we can calculate the nematic susceptibility below $T_{\rm c}$
on the bases of the linearized DW equation 
with the nematic self-energy.

\begin{figure}[htb]
\includegraphics[width=.8\linewidth]{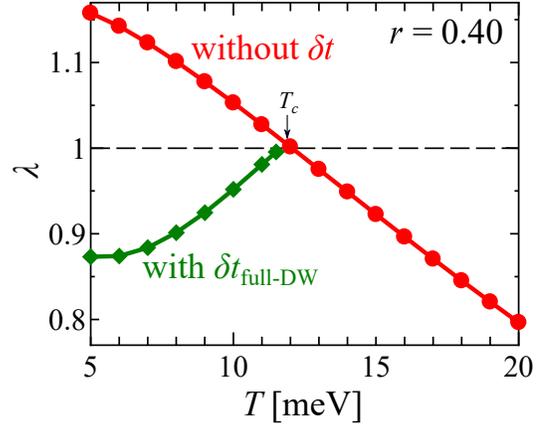}
\caption{
Eigenvalue of the DW equation in the nematic state $\lambda(T)$
with the `` nematic self-energy $\Sigma^0+\delta t$'',
where $\delta t$ is derived from the full DW equation.
Because $\Sigma^0+\delta t$ represents the thermal equilibrium state,
the nematic susceptibility $\bar{\chi}_{\rm nem}(T)=(1-\lambda(T))^{-1}$
is positive and diverges at $T=T_{\rm c}$.
}
\label{fig:heat-FeSe3}
\end{figure}

Figure \ref{fig:heat-FeSe3} 
exhibits the eigenvalue of the linearized DW equation
in the nematic state ($\Sigma\equiv \Sigma^0+\delta t$),
$\lambda_{\rm full-DW}$, in the case of $r=0.40$.
(We also show the DW equation eigenvalue with the 
normal self-energy $\Sigma^0$, $\lambda'(T)$, for reference.)
We see that $\lambda(T)$ reaches unity at $T=T_{\rm c}$,
while it monotonically decreases for $T<T_{\rm c}$.
We find that $1-\lambda(T) \approx |1-\lambda'(T)|$ for $T\le T_{\rm c}$,
as naturally expected in the GL theory.
The beautiful numerical result in Fig. \ref{fig:heat-FeSe3} 
means that the nematic-state derived from the full-DW method 
corresponds to the stationary point of $\Omega$ very accurately.
Thus, electronic states of FeSe both above and below $T_{\rm c}$
are understood in a unified way
based on the paramagnon interference mechanism.

The nematic susceptibility in the nematic state $T<T_{\rm c}$ is
$\chi_{\rm nem}(T=0)= \chi^{0f}(0)/(1-\lambda(T=0))$, 
where $1-\lambda(T=0)\approx R$.  
Since $R \ (\propto T_{\rm c}^3)$ is much smaller than unity
as shown in Fig. \ref{fig:heat-FeSe} (e),
the present study clarified that 
sizable nematic fluctuations remain in the nematic phase in Fe-based superconductors.
This is an important information to understand the pairing mechanism in FeSe.

\section{Discussions}

In this section,
we discuss important unsolved properties in Fe-based SCs
based on the present theory.
We analyze the specific heat jump at $T=T_{\rm c}$ in 
Sect. \ref{sec:Thermodynamic},
and calculate the $T$-dependence of $\chi_{\rm nem}$ 
near the nematic QCP in Sect. \ref{sec:Non-Curie-Weiss}.

\subsection{Jump in the specific heat at $T_{\rm c}$ in FeSe}
\label{sec:Thermodynamic}

From the stationary point of the free energy
 ($ \left. \d \Omega_{\rm{DW}} (T,\mu,\phi)/\d \phi \right|_{T,\mu}=0$),
we obtain
$\phi=0$ and $\phi=\sqrt{\frac{-a(T)}{b}}$ 
above and below $T_{\rm c}$, respectively.
Here, we assume a simple $T$-dependence of $a$,
$a(T)=\dot{a}(T-T_{\rm c})$, where 
$\dot{a}=-\left. \frac{d a}{d T} \right|_{T_{\rm c}} \ (>0)$.
Then, we obtain the BCS-like order parameter for $T<T_{\rm c}$ is
\begin{eqnarray}
\phi=\sqrt{\frac{\dot{a}}{b}(T_{\rm c}-T)} .
\end{eqnarray}
Then, the order parameter at $T=0$ is $\phi_0\approx\sqrt{(\dot{a}T_{\rm c})/b}$.

Based on the GL \rm{free} energy, we discuss the
jump of the heat capacity $\Delta C$ due to the DW transition,
which is calculated by
%
\begin{eqnarray}
\frac{\Delta C_{DW}}{T_{\rm c}}= -\left. \frac{d^2 \Omega_{\rm{DW}}}{d T^2}\right|_{T=T_{\rm c}}
= (\dot{a}T_{\rm c}) \left(\frac{\phi_0}{T_{\rm c}}\right)^2 .
\label{eq:GL5}
\end{eqnarray}

Now, we calculate 
$\dot{a}$ based on Eq. (\ref{eq:delsigma2}).
In FeSe, $\chi^{0f}(0)$ is almost independent of $T$,
while $ -\dot{\lambda}\equiv -\left. \frac{d\lambda}{dT}\right|_{T_{\rm c}}$
takes a large positive value as shown in Fig. \ref{fig:heat-FeSe} (d),
due to the AL-type VCs in the kernel function $I$.
Then, we obtain 
\begin{eqnarray}
(\dot{a}T_{\rm c})&=&2\chi^{0f}(0)(-\dot{\lambda}T_{\rm c}) .
\end{eqnarray}
Note that $\chi^{0f}(0)$ is equal to 
the DOS projected by the form factor $f$,
$D^{f}(0)\equiv \frac1N \sum_{k} \delta(\e_\k-\mu) f_{k}^2$,
in the absence of the self-energy.

Hereafter, we explicitly consider the mass-enhancement factor due to the self-energy,
$z^{-1}\equiv 1-\d{\rm Re}\Sigma/\d\e|_{\e=\mu}$.
The relation $z^{-1}\gg1$ holds in general strongly correlated metals.
In the Fermi liquid theory,
the Green function is given as $G_k=z/(i\e_n-z(\e_\k-\mu))$.
Then, the DOS is changed to $\chi^{0f}_z(0)= z\chi^{0f}_{z=1}(0)= zD^f(0)$, 
and the observed renormalized order parameter is 
$\phi_0^* \equiv z\phi_0$.
Thus, $\Delta C_{\rm DW}/T_{\rm c}$ due to the nematic transition is given by
\begin{eqnarray}
\frac{\Delta C_{\rm{DW}}}{T_{\rm c}}=
2z^{-1}D^{f}(0)R \left(\frac{\phi_0^*}{T_{\rm c}}\right)^2 ,
\label{eqn:DC-over-T}
\end{eqnarray}
where $R=(-\dot{\lambda}T_{\rm c})$.
As we summarized in Fig. \ref{fig:heat-FeSe} (e), $R\sim0.3$ for $r=0.40$,
and $R\sim0.1$ for $r=0.36$.
In contrast, $\Delta C_{\rm SC}/T_{\rm c}^{\rm SC}$ due to the BCS superconductivity is
${\Delta C_{\rm{SC}}}/{T_{\rm c}^{\rm SC}}=2z^{-1}D^{f}(0) \left({\psi_0^*}/{T_{\rm c}^{\rm SC}}\right)^2$
with $\Delta=\psi\cdot f$, which corresponds to $R=1$ in Eq. (\ref{eqn:DC-over-T}).
($\psi_0^* \equiv z\psi_0$ is the observed gap function.)
Because ${\psi_0^*}/{T_{\rm c}^{\rm SC}}\approx 2$,
we obtain ${\Delta C_{\rm{SC}}}/{T_{\rm c}^{\rm SC}}=8z^{-1}D^{f}(0)$,
which is close to $9.4z^{-1}D^{f}(0)$ in the BCS theory.
Because ${\phi_0^*}/{T_{\rm c}^{\rm SC}}\approx 2$ in the present numerical study, 
we obtain the relation
%
\begin{eqnarray}
\frac{\Delta C_{\rm{DW}}}{T_{\rm c}}\sim R \frac{\Delta C_{\rm SC}}{T_{\rm c}^{\rm SC}} .
\end{eqnarray}
In the present theory,
$R \propto T_{\rm c}^b$ with $b\sim3$ for $r=0.40\sim0.34$ ($T_{\rm c}=12\sim6$meV)
as shown in Fig. \ref{fig:heat-FeSe} (e).
Thus, the relation $\displaystyle \frac{\Delta C_{\rm{DW}}}{T_{\rm c}}\propto T_{\rm c}^b$
is predicted by the present theory.

\begin{figure}[htb]
\includegraphics[width=.80\linewidth]{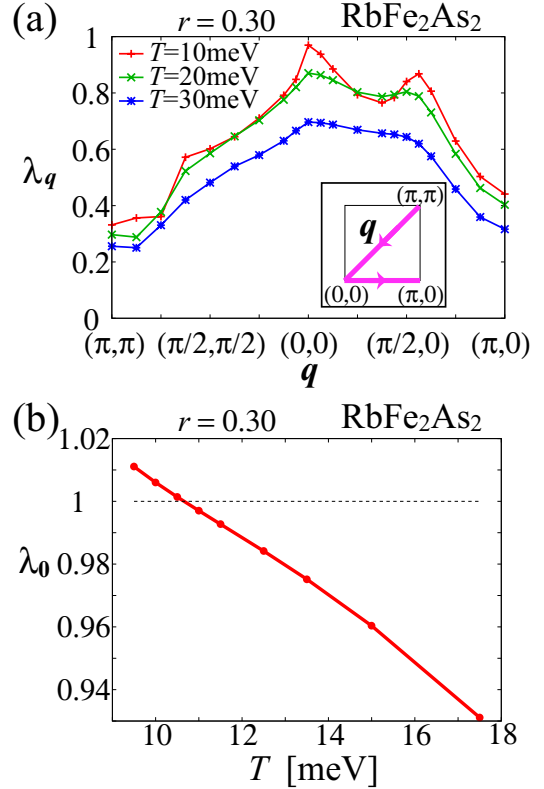}
\caption{
(a) Obtained eigenvalue $\lambda_\q$ for RbFe$_2$As$_2$ model at $r=0.30$
\cite{Onari-B2g}.
The obtained form factor at $\q={\bm0}$
is the $d_{xy}$-symmetry bond order on $xy$-orbital.
Here, we take the renormalization factor $z=1/2$
into account by following Ref. \cite{Yamakawa-PRX2016}.
(b) $T$-dependence of $\lambda_{\q={\bm0}}$ in RbFe$_2$As$_2$ model.
}
\label{fig:heat2}
\end{figure}

Next, we discuss the nematic state in RbFe$_2$As$_2$,
which is a heavily hole-doped Fe-based superconductor.
This system exhibits the uniform ($\q={\bm0}$) nematic order.
Interestingly, the observed nematicity possesses the $d_{xy}$-wave symmetry,
whose director is $45$-degrees rotated from the $d_{x^2-y^2}$-wave 
nematicity in FeSe.
Figure \ref{fig:heat2} (a) shows the DW equation eigenvalue 
in the RbFe$_2$As$_2$ Hubbard model, 
which was introduced in Ref. \cite{Onari-B2g}.
The obtained form factor at $\q={\bm0}$
is the $d_{xy}$-wave bond order on $xy$-orbital,
as we revealed in Ref. \cite{Onari-B2g}.
Here, we set $\Sigma^0=(1-z^{-1})(i\e_n-\mu)$
by following Ref. \cite{Yamakawa-PRX2016},
instead of calculating the FLEX self-energy.
We set the renormalization factor $z=1/2 \ (=m/m^*)$.
Then, $T_{\rm c}$ is renormalized to be $T_{\rm c}^*=zT_{\rm c}$,
and $\lambda_{\q}^{z=1}(T)$ is equal to $\lambda_{\q}^{z}(zT)$ 
\cite{Yamakawa-PRX2016}.

Figure \ref{fig:heat2} (b) exhibits the $T$-dependence of $\lambda_{\q={\bm0}}$.
Here, the relations $R\sim0.1$ when $T_{\rm c}\sim10$meV
and $R\propto T_{\rm c}^b$ ($b\sim2.5$) are obtained in the RbFe$_2$As$_2$ model.
This result indicates that the relation $R\sim0.01$ 
is satisfied at $T_{\rm c}\sim40$K.
In fact, in FeSe model, we obtained the relation 
$R\propto T_{\rm c}^b$ ($b\sim3$) at low $T_{\rm c}$ ($= 6\sim12{\rm meV}$)
by using fine $\k$-meshes ($64^2$) and 
many Matsubara frequencies ($8192$) 
to obtain reliable results at low $T$; see Fig. \ref{fig:heat-FeSe} (e).
Similar relation is expected to be realized in 
other Fe-based superconductor models within the same 
paramagnon interference mechanism. 
Therefore, the present theory gives a natural explanation why
$\displaystyle \frac{\Delta C_{\rm{DW}}}{T_{\rm c}}$ 
in RbFe$_2$As$_2$ ($T_{\rm c}\approx 40$K)
reported in Ref. \cite{Mizukami-B2g}
is much smaller than that in FeSe ($T_{\rm c}\approx 90$K)
in Ref. \cite{FeSe-specific-heat}.

\subsection{CW/non-CW behavior in Nematic susceptibility}
\label{sec:Non-Curie-Weiss}

\begin{figure*}[htb]
\includegraphics[width=.7\linewidth]{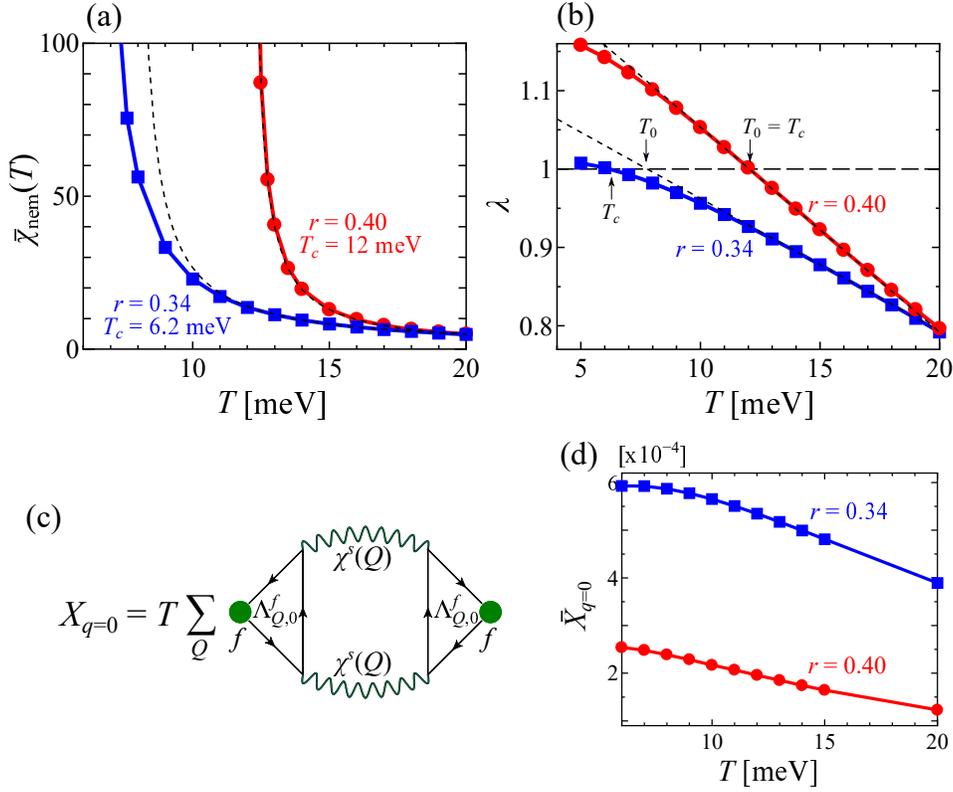}
\caption{
(a) Nematic susceptibility $\bar{\chi}_{\rm nem}=1/(1-\lambda(T))$ for 
$r=0.34$ and $0.40$ as a function of $T$.
$\bar{\chi}_{\rm nem}$ follows the CW behavior ($C/(T-T_0)$) at higher temperatures.
However, it deviates from the CW behavior at low temperatures.
(b) Derivation of the Weiss temperature $T_0$ from $\lambda(T)$.
(c) The ``paramagnon interference AL term'' 
$X_{q=0}$ that is proportional to $\lambda_{q=0}$.
$\Lambda^f_{\q={\bm0}}({\bm Q})$ is the three-point vertex.
(d) Obtained $\bar{X}_{q=0}$ as a function of $T$.
$\bar{X}_{q=0}$ starts to saturate at low temperatures, 
consistently with the deviation from the CW law in $\bar{\chi}_{\rm nem}$.
}
\label{fig:nem-suscep}
\end{figure*}

Here, we discuss the nematic susceptibility 
$\chi_{\rm nem}$ due to the electron correlation.
According to Ref. \cite{Onari-B2g,Onari-smectic},
the nematic susceptibility is given as
\begin{eqnarray}
\chi_{\rm nem}= zD^{f}(0)\frac{1}{1-\lambda(T)} ,
\end{eqnarray} 
where $\lambda(T)$ is the eigenvalue of the DW equation 
with the ``normal self-energy'' $\Sigma^0$.
Figure \ref{fig:nem-suscep} (a)
shows the normalized susceptibility
$\bar{\chi}_{\rm nem}\equiv\chi_{\rm nem}/\chi_{\rm nem}^0= 1/(1-\lambda(T))$
for $r=0.40$ and $0.34$, which corresponds to $T_{\rm c}=12$meV and 
$6.2$meV respectively.
In both cases,
$\bar{\chi}_{\rm nem}$ follows the CW behavior at higher temperatures
($T>T^*\sim 10$meV).
In contrast, at lower temperatures ($T<T^*$) for $r=0.34$, 
$\bar{\chi}_{\rm nem}$ exhibits a clear deviation from the CW behavior.
The Weiss temperature $T_0$ is derived from Fig. \ref{fig:nem-suscep} (b):
We see that $\lambda(T)$ changes from $T$-linear to $T^2$-like at 
$T\sim T^*\sim8$meV.
Similar Fermi-liquid behavior in $\lambda(T)$
is also recognized in Fig. \ref{fig:heat-FeSe} (d).
This result is natural because the system is far from the magnetic QCP.
Thus, the present theory provides a natural explanation 
for the non-CW $\chi_{\rm nem}$ near the nematic 
QCP with $T_{\rm c}\approx 0$ reported in Refs. \cite{Hosoi}.

Here, we discuss the reason why $\chi_{\rm nem}$
exhibits the CW/non-CW behavior depending on $r$.
According to Eq. (\ref{eq:lambda_q}), 
the eigenvalue $\lambda_q$ at $q=0$ is proportional to $X$ in Eq. (\ref{eq:X_q})
because $\chi^{0f}(q=0)$ is almost constant.
In FeSe,
the nematic state is mainly caused by the
``paramagnon interference AL term'' in $X_q$.
It is approximately given as
\begin{eqnarray}
X_{q=0}&=&T\sum_{Q,m} 3|\Lambda_{Q,{q=0}}^{f,m}|^2 (U^2\chi^{s,m}_Q)^2 ,
\label{eqn:Xq} 
\end{eqnarray}
%
where $\chi^{s,m}_Q$ is the spin susceptibility for $d$-orbital $m$,
and $\Lambda_{Q,q}^{f,m}$ is the three-point vertex 
composed of three Green functions:
\begin{eqnarray}
\Lambda_{Q,{q=0}}^{f,m} &=& -T\sum_{p}
({\hat G}^0_p {\hat f}_{p}^{q=0} {\hat G}^0_{p})_{m,m}({\hat G}^0_{p+Q})_{m,m}\} ,
\label{eqn:LQq}
\end{eqnarray}
where ${\hat G}^0_p$ is the matrix representation of the 
multiorbital Green function with FLEX self-energy.
Their expressions are shown in Fig. \ref{fig:nem-suscep} (c).
In a simple single-orbital model,
the analytic expression of Eq. (\ref{eqn:LQq}) is given as
$\displaystyle \Lambda_{Q,q=0}^f= \frac1N \sum_\k \left(-\frac{\d n(\e_\k)}{\d \e_\k}\right)\frac{1}{\e_{\k-\Q}-\e_\k} f_\k^{q=0}$,
where $n(\e)$ is the Fermi distribution function \cite{Kontani-Raman}.
When $\Q\approx\Q_{\rm nesting}$, $\Lambda_{Q,q=0}^f$ exhibits 
strong enhancement at low temperatures due to $(-\d n(\e_\k)/\d\e_\k)$
\cite{Yamakawa-PRX2016,Kontani-Raman}.

In FeSe, the spin Stoner factor $\a_S$ is nearly constant for $T>T_{\rm c}$.
Then, the strong $T$-dependence of $\lambda(T)$ originates from 
$\Lambda_{Q,{q=0}}^f$, not from $V^s_Q$, as we explained
in Ref. \cite{Yamakawa-PRX2016}.
The paramagnon interference magnifies the nematic susceptibility,
but its magnification is nearly constant. 
To confirm this fact, we introduce a simplification of $X_{q=0}$
in Eq. (\ref{eqn:Xq}) as follows:
\begin{eqnarray}
\bar{X}_{q=0}&=&T\sum_{Q,m} |\Lambda_{Q,{q=0}}^{f,m}|^2 .
\label{eqn:Xq2}
\end{eqnarray}
Figure \ref{fig:nem-suscep} (d) shows the 
numerical result of $\bar{X}_{q=0}$ at $r=0.34$.
For $T\gtrsim10$meV,
$\bar{X}_{q=0}$ exhibit almost perfect $T$-linear behavior.
Its increment at low $T$ originates from the FS nesting.
In contrast, at lower temperatures, $\bar{X}_{q=0}$ starts to saturate 
when $T$ is smaller than the nesting energy scale 
\cite{Yamakawa-PRX2016}.
This saturation gives rise to the non-CW behavior of 
$\chi_{\rm nem}$ near the nematic QCP ($T_{\rm c}\sim0$)
as shown in Fig. \ref{fig:nem-suscep} (a).

Thus, the paramagnon interference mechanism satisfactorily explains
both the CW behavior of $\chi_{\rm nem}$ above $T^*(\sim10{\rm meV})$
and its non-CW behavior below $T^*$.
These behaviors are actually observed in various Fe-based superconductors 
near the nematic QCP: Ba(Fe,T)$_2$As$_2$ with T=Co,Ni, (Ba,A)Fe$_2$As$_2$ with A=K,Rb, 
and Fe(Se,Pn) with Pn=Te,S
\cite{Fisher-Science2016,Shibauchi-nemQCP,Terashima}.
(In this mechanism, $\Lambda_{Q,{q=0}}^f$ is the coupling constant 
between the nematicity and the paramagnons,
and its increment leads to large $\chi_{\rm nem}$ at low temperatures.)
Once the nematic order is established below $T_c$, the spin Stoner factor $\a_S$ increases as we discussed in \cite{Onari-FeSe,saigo}.
\color{black}

In the present mechanism, 
the deviation from the CW behavior of $\chi_{\rm nem}$ below $T^*$
is equal to the Fermi-liquid behavior $\lambda(0)-\lambda(T)\propto T^2$.
This deviation is naturally expected when the nematic QCP is well 
separated from the magnetic QCP, even in the absence of impurities.
We will discuss this point in the Summary section.

\section{Summary}

In this paper,
we derived a formally exact ``density-wave (DW) equation'',
by introducing the form factor of the DW state $\delta t^{\q\s}_{\k}$
into the LW theory.
Its solution automatically satisfies the extremum condition of the grand potential. 
By solving the DW equation, the optimized form factor and its wavevector
are uniquely obtained for both above and below $T_{\rm c}$.
This formalism enables us to perform 
the Baym-Kadanoff conserving approximation
that is essential to obtain thermodynamic stable states.
In addition, we derive an exact expression of the 
Ginzburg-Landau (GL) free energy, $F\propto a_\q\phi^2$,
where $\phi$ is the amplitude of the DW order at wavevector $\q$.
The coefficient 
$a_\q \ [\approx a_{\q_0}+\frac12\sum_{\mu,\nu}c_{\mu,\nu}(q^\mu-q_0^\mu)(q^\nu-q_0^\nu)]$
is uniquely related to the eigenvalue of the DW equation $\lambda_\q$.
This formalism enables us to calculate various 
thermodynamic properties of the DW state.

In the second part,
we analyzed the nematic state in FeSe based on the derived DW equation
based on a realistic multiorbital Hubbard model
with one single parameter $r$.
We explained the following key experiments in Fe-based SCs:
{\bf (i)} Lifshitz transition due to bond+orbital order
\cite{FeSe-Lif1,FeSe-Lif2}.
{\bf (ii-1)} The CW behavior of $\chi_{\rm nem}\propto 1/|1-\lambda(T)|$ 
at higher-temperatures; $1-\lambda(T)\propto T_0-T$.
{\bf (ii-2)} Deviation from the CW behavior of $\chi_{\rm nem}$ 
at low temperatures near the nematic QCP without magnetic criticality;
$\lambda(0)-\lambda(T)\propto T^2$
\cite{Hosoi,B2g-Ishida,Terashima}.
{\bf (iii)} A scaling relation $\Delta C/T_{\rm c} \propto T_{\rm c}^b$ ($b\sim3$)
that naturally explains the smallness of $\Delta C/T_{\rm c}$
reported in several nematic systems
\cite{Sato-CDW,Murayama-CDW,Mizukami-B2g}.
This is because the gain of the free energy in the nematic transition
is much smaller than that in the SC state.
In addition, we explain the
{\bf (iv)} Nematic QCP away from the magnetic QCP
observed in Fe(Se,S), Fe(Se,Te), \cite{Shibauchi-nemQCP}
and Na(Fe,Co)As \cite{Zheng-NaFeAs}.

The present theory naturally explains the 
essential points {\bf (i)}-{\bf (iv)}.
Thus, it is concluded that the nematicity in FeSe is the bond+orbital order
due to the ``paramagnon interference mechanism''
depicted in Fig. \ref{fig:nem-suscep} (c)
\cite{Onari-SCVC,Yamakawa-PRX2016,Onari-FeSe,Tazai-JPSJ-fRG}.

The behavior {\bf (ii-2)} 
has been observed in various Fe-based superconductors near the nematic QCP: 
Ba(Fe,T)$_2$As$_2$ with T=Co,Ni, (Ba,A)Fe$_2$As$_2$ with A=K,Rb, and Fe(Se,Pn) with Pn=Te,S
\cite{Fisher-Science2016,Shibauchi-nemQCP}.
This behavior is frequently ascribed to the impurity-induced Griffiths phase, 
while it is widely observed insensitively to the impurity potential strength.
(For example, the quantum oscillation is observed in Te- and S-doped FeSe.)
In the present theory,
the behavior {\bf (ii-2)} is naturally explained when the nematic QCP is 
well separated from the magnetic QCP, even in the absence of impurities.
It is useful to verify the relation 
$\lambda(0)-\lambda(T)\propto T^2$ experimentally.
In the present mechanism,
the increment of $\chi_{\rm nem}$ at low $T$
originates from the $T$-dependence of $\Lambda_{Q,{\q=\bm{0}}}$
in Eq. (\ref{eqn:LQq}) 
\cite{Yamakawa-PRX2016},
and the self-energy due to thermal spin fluctuations
\cite{Kontani-ROP,Onari-transport-FeSe}
is also important to derive a perfect CW behavior of $\chi_{\rm nem}$.
The self-energy due to nematic fluctuations 
will be also important ($|\Sigma_{\rm nem}|\gtrsim|\Sigma_{\rm FLEX}|$)
adjacent to the nematic QCP, unless the dynamical nematic fluctuations 
are suppressed by the acoustic phonons.
This is an important future issue.
It is considered that a perfect CW behavior for $30$-$250$K observed 
in BaFe$_2$(As$_{0.3}$P$_{0.7}$)$_2$
\cite{Fisher-Science2016}
is ascribed to the magnetic criticality 
due to the coincidence of the nematic and magnetic QCPs.

The present theory paves the way for understanding
various unconventional phase transition systems
for both above and below $T_{\rm c}$.
For example, the analysis of the odd-parity DW order 
accompanying spontaneous current,
which has been reported in cuprates and kagome metals recently,
is an important future problem
\cite{Tazai-cLC,Kontani-sLC}.
The (local and/or non-local) multipole order physics 
in $5d$- and $f$-electron systems 
with strong spin-orbit interaction is another important future issue
\cite{Tazai-CeB6}.
In addition, it is important to develop the numerical method beyond the one-loop approximation.
the functional-renormalization-group method
\cite{Metzner-RMP,Honerkamp,Tsuchiizu-PRL2013,Tazai-cLC,Tazai-JPSJ-fRG},
which is equivalent to the parquet equation,
would be useful to obtain a reliable kernel function of the DW equation.
The exotic superconductivity mediated by the DW fluctuations
\cite{Kontani-PRL2010,Tazai-kagome,Yamakawa-FeSe-underP,Yamakawa-FeSe-SC}
would be a very interesting future problem.

\acknowledgements
We are grateful to E.-G. Moon for very enlightening discussions.
We also thank Y. Matsuda, T, Shibauchi, T. Hashimoto, and Y. Mizukami
for fruitful discussions on experiments.
This work is supported by Grants-in-Aid for Scientific Research (KAKENI) 
Research (No. JP20K22328, No. JP20K03858, No. JP19H05825, No. JP18H01175) 
from MEXT of Japan.

\appendix

\section{Justification of Onari-Kontani approximation in the DW equation}

\begin{figure}[htb]
\includegraphics[width=.98\linewidth]{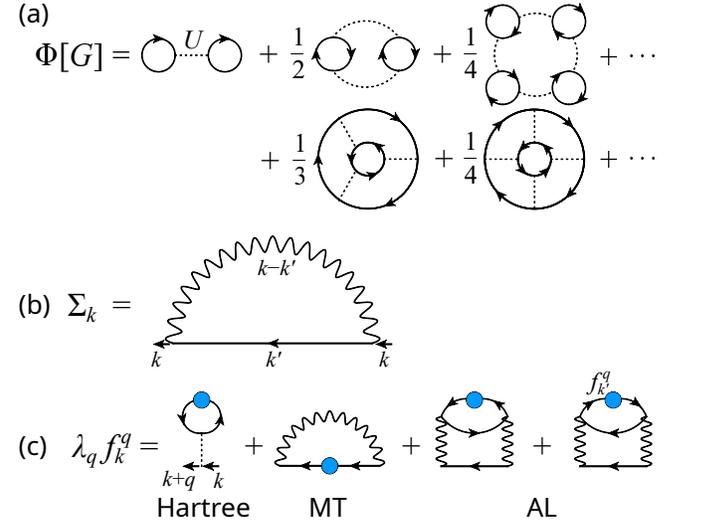}
\caption{
(a) Diagrammatic expressions of $\Phi_{\rm FLEX}[G]$ in the one-loop approximation.
For simplicity, diagrams in the single-orbital Hubbard models are shown,
while we study multi-orbital Hubbard models in this paper.
(b) Self-energy $\Sigma_k=\delta\Phi_{\rm FLEX}/\delta G_k$.
(c) Linearized DW equation with the kernel function $I$
derived from the second derivative of $\Phi_{\rm FLEX}[G]$ with respect to $G$.
The Maki-Thompson (MT) term and the 
Aslamazov-Larkin (AL) terms give the 
``fluctuation-induced interaction for the DW''.
}
\label{fig:diagram}
\end{figure}

Here, we derive the DW equation introduced
by Onari and Kontani {\it et al.} in 
Refs. \cite{Onari-FeSe,Kawaguchi-CDW,Tsuchiizu-CDW,Tazai-JPSJ-fRG},
which has been applied to iron-based and cuprate superconductors, 
from the exact DW equation given in Eq. (\ref{eq:cons06}).
In the present calculation, 
we apply the fluctuation-exchange (FLEX) approximation 
for $\Phi_{\rm{FLEX}}[G]$.
It is given as \cite{Yanase}
\begin{eqnarray}
\Phi_{\rm FLEX}&=&T\sum_q{\rm Tr}\left\{ \frac32\ln(1-U^s\chi^0_q)
+\frac12\ln(1-U^c\chi^0_q)\right\}
\nonumber \\
& &+\frac{T}{4}\sum_q {\rm Tr}\left\{(U^s\chi^0_q)^2+(U^c\chi^0_q)^2\right\}
\nonumber \\
& &+T\sum_q {\rm Tr}\left\{\frac32 U^s\chi^0_q+\frac12 U^c\chi^0_q\right\} ,
\end{eqnarray}
which is expressed in Fig.\ref{fig:diagram} (a).
Here, $U^{s(c)}$ is the spin-channel (charge-channel) Coulomb interaction,
and $U^s=-U^c=U$ in the single-orbital Hubbard model.
Their matrix expressions in multiorbital systems
are introduced in the next section.

In the framework of the conserving approximation,
the first-order derivative of $\Phi_{\rm{FLEX}}[G]$
gives the self-energy $\Sigma$.
It is expressed as \cite{Bickers,Onari-SCVCS}
\begin{eqnarray}
\Sigma^0_{k}=\frac{T}{N}\sum_q G^0_{k+q}W^0_{q} ,
\label{eqn:self0}
\end{eqnarray}
which is expressed in Fig.\ref{fig:diagram} (b).
Here, $\displaystyle W^0_{q}=\left(\frac32 V^s_{q}+\frac12 V^c_{q}\right)$,
$V^x_{q}= U^x+U^x\chi^x_{q}U^x$ ($x=s,c$),
and $\chi^{x}_q=\chi^{0}(q)/(1-U^{x}\chi^{0}(q))$.

Finally, we derive the irreducible four-point vertex $I$
from the second derive of $\Phi_{\rm FLEX}[G]$.
The derived charge-channel kernel function in the DW equation
(\ref{eq:cons06cs}) for $x=c$ is given by
\cite{Kawaguchi-CDW,Kontani-sLC}
\begin{eqnarray}
& &\!\!\!\!\!\!\!\!\!\!
I^c_{kk'q}=
-\frac{3}{2} V^{s}_{k-k'}-\frac{1}{2} V^{c}_{k-k'} 
\nonumber \\
& & 
+\frac{T}{N}\sum_{p}
 [\frac{3}{2} V^{s}_{p+q}V^{s}_{p}+\frac{1}{2} V^{c}_{p+q}V^{c}_{p}]G^0_{k-p}G^0_{k'-p}
\nonumber \\
& & 
+\frac{T}{N}\sum_{p}
 [\frac{3}{2} V^{s}_{p+q}V^{s}_{p}+\frac{1}{2} V^{c}_{p+q}V^{c}_{p}]G^0_{k-p}G^0_{k'+p} ,
\label{eqn:IcS} 
\end{eqnarray}
which is depicted in Fig.\ref{fig:diagram} (c).
Note that the double-counting $U^2$-terms in Eqs. (\ref{eqn:IcS})
and (\ref{eqn:self0}) should be subtracted properly.

Although the DW equation with the kernel in Eq. (\ref{eqn:IcS}) 
is an approximation, it satisfies the Baym-Kadanoff's conserving laws
by introducing $\Sigma_{\rm FLEX}$.
That is, the solution of the DW equation is
the thermal equilibrium state
derived from the stationary condition of $\Omega_{\rm{FLEX}}$.
Thus, the Onari-Kontani type DW equation
\cite{Onari-FeSe,Kawaguchi-CDW,Tsuchiizu-CDW,Tazai-JPSJ-fRG} 
is given by dropping $\Sigma_{\rm FLEX}$ from Eq. (\ref{eqn:IcS}).
Furthermore, the present exact DW equation is useful to 
go beyond the Onari-Kontani's approximation.
\section{GL free energy from $\Omega[G]$}

In the main text, we derive the GL free energy based on
the grand potential $\Omega[\Sigma]$.
Here, we show a different way to obtain GL free energy by using $\Omega[G]$
starting from Eq. (\ref{eq:omega0}) in the main text.
%
%
The coefficient $a$ is derived from the second functional derivation of $\Omega[G]$ as
\begin{eqnarray}
\Omega[\bar{G}]-\Omega[G^0] &=&
\sum_{k\sigma k'\sigma'}  \left. \frac{\dd^2 \Omega[G]}{\dd G_{k'\sigma'}\dd G_{k\sigma}}\right|_{G=G^0} 
\nonumber \\
& &\times \dd G_{k'\sigma'} \dd G_{k\sigma} \\
\frac{\dd^2 \Omega[G]}{\dd G_{k'\sigma'}\dd G_{k\sigma} }&=&
 \frac{\dd}{\dd G_{k'\sigma'}}\left\{ G_{k\sigma}^{-1} -(G^{\rm{\rm{free}}}_{k\sigma})^{-1}+\Sigma_{k\sigma}[G] \right\} \nonumber \\
&=& -G_{k\sigma}^{-2} \delta_{kk'}-\frac{\dd \Sigma_{k\sigma}[G]}{\dd G_{k'\sigma'}} 
\nonumber \\
\left. \frac{\dd^2 \Omega[G]}{\dd G_{k\sigma}\dd G_{k'\sigma'} } \right| _{G=G^0}
&=& -(G^0_{k\sigma})^{-2}\delta_{kk'}- I_{kk'}^{\sigma \sigma'} .
\end{eqnarray} 
Thus, we obtain the following results
\begin{eqnarray}
\Omega[G]-\Omega[G^0] &=&
\frac{T}{N}\sum_{k\sigma} (G^0_{k\sigma})^{-2}(\dd G_{k\sigma})^2 
\nonumber \\
& &- \frac{T}{N}\sum_{k\sigma k'\sigma'} 
I_{kk'}^{\sigma \sigma'} \dd G_{k\sigma} \dd G_{k'\sigma'} 
\nonumber \\
&=& \chi^{f}(0) \left\{ 1-\lambda \right\} \phi^2,
\label{eq:omega_tenkai4-S}
\end{eqnarray}
where we use the relation $\dd G_{k\sigma}=(G^0_{k\sigma})^2\delta t_{k\sigma}$
and set $t_{k\sigma}=f_{k\sigma}\cdot \phi$.
Therefore, the coefficient $a$ of GL free energy from $\Omega[G]$ is given by
\begin{eqnarray}
a&=&  \chi^{f}(0) \left\{ 1-\lambda \right\} \phi^2 .
\label{machi1}
\end{eqnarray}
At $T=T_{\rm c}$, the obtained $a$ by $\Omega[G]$ is the same as the one by $\Omega[\Sigma]$ in the main text, while it becomes different at $T\neq T_{\rm c}$.
Moreover, the expression $a$ by $\Omega[G]$ does not reproduce the results by mean-field approximation. Thus, the expression of Eq. (\ref{machi1}) is correct only at $T=T_{\rm c}$

\section{Eight-orbital models for FeSe}

\begin{figure}[htb]
\includegraphics[width=.8\linewidth]{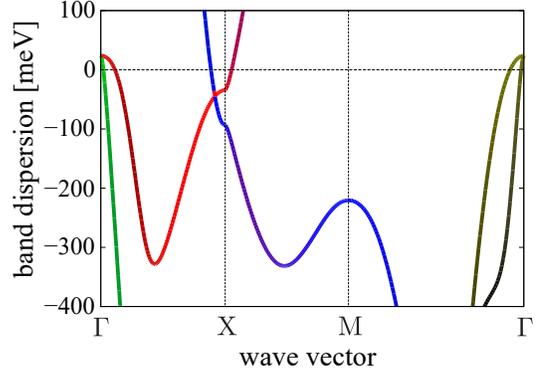}
\caption{
Band-dispersion of the present FeSe model without self-energy.
The renormalized dispersion due to the self-energy 
is shown in Fig. \ref{fig:heat-FeSe2}.
}
\label{fig:band-original}
\end{figure}

Here, we introduce the eight-orbital $d$-$p$ models for FeSe. 
We first derived the first-principles tight-binding models
using the WIEN2k and WANNIER90 codes.
Crystal structure parameters of FeSe 
is given in Refs. \cite{FeSe-t}.
By following Ref. \cite{Yamakawa-PRX2016},
we introduce the $k$-dependent shifts for orbital $l$, $\delta E_l$,
in order to obtain the experimentally observed FSs
\cite{FeSe-Lif1,FeSe-Lif2}.
In this paper, we introduce the intra-orbital hopping parameters into 
the first-principles FeSe model 
in order to shift the $d_{xy}$ orbital band ($d_{xz/yz}$-orbital band) at 
($\Gamma$, M, X) points by (0,-0.35,+0.40) [(-0.22,0,+0.16)] 
in units of eV. 
Such level shifts are introduced by the additional intra-orbital
 \cite{Yamakawa-PRX2016}.
The band-dispersion of the present FeSe model
without self-energy is shown in Fig. \ref{fig:band-original}.
Its Fermi surface is given in 
Fig. \ref{fig:heat-FeSe} (a) in the main text.

In this multiorbital model, 
the matrix expression of the non-interacting Green function 
is given as
\begin{eqnarray}
{\hat G}_k^{\rm free} = ((\e_n-\mu){\hat 1}+{\hat h}_\k^0)^{-1} ,
\label{eqn:Gfree-S}
\end{eqnarray}
where $k\equiv (\k,\e_n)$,
and ${\hat h}_\k^0$ is the matrix expression 
of the kinetic term, which is given by the 
Fourier transformation of the tight-binding model.

We also explain the multiorbital Coulomb interaction term $H_U$.
The multiorbital Coulomb interaction term is expressed as
$H_U= \frac14 \sum_{i,\s\s'} \sum_{ll'mm'} U^{\s\s'}_{ll'mm'} d_{i,l,\s}^\dagger d_{i,m',\s'}^\dagger d_{i,m,\s'} d_{i,l',\s}$,
where $l,m$ represent the orbital indices, in $\s=+1 \ (-1)$ represents the 
$\uparrow \ (\downarrow)$ spin, $i$ is the site index,
and ${\hat U}^{\s\s'}=-{\hat U}^c-\s\s'{\hat U}^s$.
The matrix elements of ${\hat U}^s$ is given by
\color{black}
\cite{Yamakawa-PRX2016}
\begin{equation}
({\hat U}^{\mathrm{s}})_{l_{1}l_{2},l_{3}l_{4}} = \begin{cases}
U_{l_1,l_1}, & l_1=l_2=l_3=l_4 \\
U_{l_1,l_2}' , & l_1=l_3 \neq l_2=l_4 \\
J_{l_1,l_3}, & l_1=l_2 \neq l_3=l_4 \\
J_{l_1,l_2}, & l_1=l_4 \neq l_2=l_3 \\
0 , & \mathrm{otherwise}.
\end{cases}
\label{eqn:Us-S}
\end{equation}
Also, the bare Coulomb interaction for the charge channel is
\begin{equation}
({\hat U}^{\mathrm{c}})_{l_{1}l_{2},l_{3}l_{4}} = \begin{cases}
-U_{l_1,l_1}, & l_1=l_2=l_3=l_4 \\
U_{l_1,l_2}'-2J_{l_1,l_2} , & l_1=l_3 \neq l_2=l_4 \\
-2U_{l_1,l_3}' + J_{l_1,l_3} , & l_1=l_2 \neq l_3=l_4 \\
-J_{l_1,l_2} , &l_1=l_4 \neq l_2=l_3 \\
0 . & \mathrm{otherwise}.
\end{cases}
\label{eqn:Uc-S}
\end{equation}
Here, $U_{l,l}$, $U_{l,l'}'$ and $J_{l,l'}$
are the first-principles Coulomb interaction terms
given in Ref. \cite{Miyake}.

In the main text, 
we omit the orbital indices of the Green functions 
and the Coulomb interactions to simplify the expressions.
It is straightforward to write the orbital indices
of these expressions in multiorbital models
by using Eqs. (\ref{eqn:Gfree-S})-(\ref{eqn:Uc-S}).

\section{Relative phase between $xz$-orbital and $xy$-orbital form factors}

As we discussed in the main text,
the relation $2|\delta t_{xz}^*|\approx|\delta t_{xy}^*|$
in Fig. \ref{fig:heat-FeSe2} (a) 
indicates that both the $(xz+yz)$-orbitals and $xy$-orbital
equally contribute to the nematic order in FeSe.
The $(xz+yz)$-orbital polarization [$xy$-orbital bond order]
originates from the spin fluctuations on 
$(xz+yz)$-orbitals [$xy$-orbital].
Here, we discuss the relative phase between 
$xz$-orbital and $xy$-orbital form factors,
$\delta t_{xz,xz}(0,\pi)\times\delta t_{xy,xy}(0,\pi)<0$,
based on the Ginzburg-Landau analysis.

According to Eq. (\ref{eq:omega_tenkai3-2}) or (\ref{eq:omega_tenkai4}),
the second-order inter-orbital free-energy is
$F^{(2)}_{2,4}= -2\chi^{0f}_{2,4}\phi_2\phi_4$, where 
$\chi_{2,4}^{0\delta t}=-T\sum_kG_{2,4}(k)G_{4,2}(k)f_{2,2}(k)f_{4,4}(k)$
and $\delta t_{m,m}(k)=f_{m,m}(k)\phi_m$.
We verified numerically that $\chi_{2,4}^{0f}>0$,
which is consistent with the relations $\chi^0_{22,44}(0)<0$ in FeSe
and $\delta t_{2,2}(k)\delta t_{4,4}(k)<0$ around Y-point
shown in Fig. \ref{fig:heat-FeSe} (c).
Thus, the relation $\phi_2\phi_4>0$ is realized,
and therefore the Lifshitz transition occurs in FeSe.
In other words, the relation $\phi_2\phi_4>0$ 
({i.e.}, $\delta t_{2,2}(0,\pi)\delta t_{4,4}(0,\pi)<0$)
is a direct consequence of the multiorbital band structure 
of Fe-based superconductors.




\vspace{10mm}

\begin{thebibliography}{999}
\bibitem{Chubukov-nematic-rev}
A. V. Chubukov, M. Khodas, and R. M. Fernandes,
{\it Magnetism, Superconductivity, and Spontaneous Orbital Order in Iron-Based Superconductors: Which Comes First and Why?},
Phys. Rev. X {\bf 6}, 041045 (2016).

\bibitem{Yamakawa-PRX2016}
Y. Yamakawa, S. Onari, and H. Kontani,
{\it Nematicity and Magnetism in FeSe and Other Families of Fe-Based Superconductors},
Phys. Rev. X {\bf 6}, 021032 (2016).

\bibitem{Onari-FeSe}
S. Onari, Y. Yamakawa, and H. Kontani,
{\it Sign-Reversing Orbital Polarization in the Nematic Phase of FeSe due to the ${C}_{2}$ Symmetry Breaking in the Self-Energy},
Phys. Rev. Lett. {\bf 116}, 227001 (2016).

\bibitem{Fernandes-Nature}
R. M. Fernandes, A. V. Chubukov, and J. Schmalian,
{\it What drives nematic order in iron-based superconductors?},
Nat. Phys. {\bf 10}, 97--104 (2014).

\bibitem{Yoshizawa}
M. Yoshizawa, D. Kimura, T. Chiba, S. Simayi, Y. Nakanishi, K. Kihou, C.-H. Lee, A. Iyo, H. Eisaki, M. Nakajima, and S.-i. Uchida,
{\it Structural Quantum Criticality and Superconductivity in Iron-Based Superconductor Ba(Fe$_{1-x}$Co$_x$)$_2$As$_2$},
J. Phys. Soc. Jpn. {\bf 81}, 024604 (2012).

\bibitem{Fernandes1}
R. M. Fernandes, L. H. VanBebber, S. Bhattacharya, P. Chandra, V. Keppens, D. Mandrus, M. A. McGuire, B. C. Sales, A. S. Sefat, and J. Schmalian,
{\it Effects of Nematic Fluctuations on the Elastic Properties of Iron Arsenide Superconductors},
Phys. Rev. Lett. {\bf 105}, 157003 (2010).

\bibitem{Bohmer}
A. E. B\"ohmer, P. Burger, F. Hardy, T. Wolf, P. Schweiss, R. Fromknecht, M. Reinecker, W. Schranz, and C. Meingast,
{\it Nematic Susceptibility of Hole-Doped and Electron-Doped ${\mathrm{Ba}\mathrm{F}\mathrm{e}}_{2}{\mathrm{As}}_{2}$ Iron-Based Superconductors from Shear Modulus Measurements},
Phys. Rev. Lett. {\bf 112}, 047001 (2014).

\bibitem{Gallais}
Y. Gallais, R. M. Fernandes, I. Paul, L. Chauvi\`ere, Y.-X. Yang, M.-A. M\'easson, M. Cazayous, A. Sacuto, D. Colson, and A. Forget,
{\it Observation of Incipient Charge Nematicity in ${\mathrm{Ba}(\mathrm{Fe}{}_{1\ensuremath{-}X}\mathrm{Co}{}_{X})}_{2}\mathrm{As}{}_{2}$},
Phys. Rev. Lett. {\bf 111}, 267001 (2013).

\bibitem{Gallais2}
P. Massat, D. Farina, I. Paul, S. Karlsson, P. Strobel, P. Toulemonde, M.-A. M\ifmmode \mbox{\'{e}}\else \'{e}\fi{}asson, M. Cazayous, A. Sacuto, S. Kasahara, T. Shibauchi, Y. Matsuda, and Y. Gallais,
{\it Charge-induced nematicity in FeSe},
Proc. Natl. Acad. Sci. U.S.A. {\bf 113}, 9177 (2016).

\bibitem{Raman-spectroscopy}
T. Kissikov, R. Sarkar, M. Lawson, B. T. Bush, E. I. Timmons, M. A. Tanatar, R. Prozorov, S. L. Bud'ko, P. C. Canfield, R. M. Fernandes, and N. J. Curro,
{\it Uniaxial strain control of spin-polarization in multicomponent nematic order of BaFe2As2},
Nat. Commun. {\bf 9}, 1058 (2018).

\bibitem{Fisher-review}
I. R. Fisher, L. Degiorgi, and Z. X. Shen,
{\it In-plane electronic anisotropy of underdoped `122' Fe-arsenide superconductors revealed by measurements of detwinned single crystals},
Rep. Prog. Phys. {\bf 74}, 124506 (2011).

\bibitem{Hosoi}
S. Hosoi, K. Matsuura, K. Ishida, H. Wang, Y. Mizukami, T. Watashige, S. Kasahara, Y. Matsuda, and T. Shibauchi,
{\it Nematic quantum critical point without magnetism in FeSe$_{1\&\#x2212;x}$S$_{x}$ superconductors},
Proc. Natl. Acad. Sci. U.S.A. {\bf 113}, 8139 (2016).

\bibitem{B2g-Ishida}
K. Ishida, M. Tsujii, S. Hosoi, Y. Mizukami, S. Ishida, A. Iyo, H. Eisaki, T. Wolf, K. Grube, H. v. L\ifmmode \mbox{\"{o}}\else \"{o}\fi{}hneysen, R. M. Fernandes, and T. Shibauchi,
{\it Novel electronic nematicity in heavily hole-doped iron pnictide superconductors},
Proc. Natl. Acad. Sci. U.S.A. {\bf 117}, 6424 (2020).

\bibitem{Fisher-Science2016}
H.-H. Kuo, J.-H. Chu, J. C. Palmstrom, S. A. Kivelson, and I. R. Fisher,
{\it Ubiquitous signatures of nematic quantum criticality in optimally doped Fe-based superconductors},
Science {\bf 352}, 958 (2016).

\bibitem{Fernandes-TBG}
R. M. Fernandes and J. W. F. Venderbos,
{\it Nematicity with a twist: Rotational symmetry breaking in a moir\&\#xe9; superlattice},
Sci. Adv. {\bf 6}, eaba8834 (2020).

\bibitem{Onari-TBG}
S. Onari and H. Kontani,
{\it SU(4) $\text{Valley}+\text{Spin}$ Fluctuation Interference Mechanism for Nematic Order in Magic-Angle Twisted Bilayer Graphene: The Impact of Vertex Corrections},
Phys. Rev. Lett. {\bf 128}, 066401 (2022).

\bibitem{Nakaoka-Ti}
H. Nakaoka, Y. Yamakawa, and H. Kontani,
{\it Theoretical prediction of nematic orbital-ordered state in the Ti oxypnictide superconductor ${\mathrm{BaTi}}_{2}{(\mathrm{As},\mathrm{Sb})}_{2}\mathrm{O}$},
Phys. Rev. B {\bf 93}, 245122 (2016).

\bibitem{Sato-CDW}
Y. Sato, S. Kasahara, H. Murayama, Y. Kasahara, E.-G. Moon, T. Nishizaki, T. Loew, J. Porras, B. Keimer, T. Shibauchi, and Y. Matsuda,
{\it Thermodynamic evidence for a nematic phase transition at the onset of the pseudogap in YBa2Cu3Oy},
Nat. Phys. {\bf 13}, 1074--1078 (2017).

\bibitem{Murayama-CDW}
H. Murayama, Y. Sato, R. Kurihara, S. Kasahara, Y. Mizukami, Y. Kasahara, H. Uchiyama, A. Yamamoto, E.-G. Moon, J. Cai, J. Freyermuth, M. Greven, T. Shibauchi, and Y. Matsuda,
{\it Diagonal nematicity in the pseudogap phase of HgBa2CuO4+\ifmmode {\delta}\else ${\delta}$\fi{}},
Nat. Commun. {\bf 10}, 3282 (2019).

\bibitem{Kawaguchi-CDW}
K. Kawaguchi, Y. Yamakawa, M. Tsuchiizu, and H. Kontani,
{\it Competing Unconventional Charge-Density-Wave States in Cuprate Superconductors: Spin-Fluctuation-Driven Mechanism},
J. Phys. Soc. Jpn. {\bf 86}, 063707 (2017).

\bibitem{Tsuchiizu-CDW}
M. Tsuchiizu, K. Kawaguchi, Y. Yamakawa, and H. Kontani,
{\it Multistage electronic nematic transitions in cuprate superconductors: A functional-renormalization-group analysis},
Phys. Rev. B {\bf 97}, 165131 (2018).

\bibitem{Tazai-kagome}
R. Tazai, Y. Yamakawa, S. Onari, and H. Kontani,
{\it Mechanism of exotic density-wave and beyond-Migdal unconventional superconductivity in kagome metal AV$_{3}$Sb$_{5}$ (A = K, Rb, Cs)},
Sci. Adv. {\bf 8}, eabl4108 (2022).

\bibitem{kagome-Thomale2013}
M. L. Kiesel, C. Platt, and R. Thomale,
{\it Unconventional Fermi Surface Instabilities in the Kagome Hubbard Model},
Phys. Rev. Lett. {\bf 110}, 126405 (2013).

\bibitem{kagome-SMFRG}
W.-S. Wang, Z.-Z. Li, Y.-Y. Xiang, and Q.-H. Wang,
{\it Competing electronic orders on kagome lattices at van Hove filling},
Phys. Rev. B {\bf 87}, 115135 (2013).

\bibitem{kagome-Balents2021}
T. Park, M. Ye, and L. Balents,
{\it Electronic instabilities of kagome metals: Saddle points and Landau theory},
Phys. Rev. B {\bf 104}, 035142 (2021).

\bibitem{Yamakawa-CDW}
Y. Yamakawa and H. Kontani,
{\it Spin-Fluctuation-Driven Nematic Charge-Density Wave in Cuprate Superconductors: Impact of Aslamazov-Larkin Vertex Corrections},
Phys. Rev. Lett. {\bf 114}, 257001 (2015).

\bibitem{Sachdev-CDW1}
S. Sachdev and R. La Placa,
{\it Bond Order in Two-Dimensional Metals with Antiferromagnetic Exchange Interactions},
Phys. Rev. Lett. {\bf 111}, 027202 (2013).

\bibitem{Husemann}
C. Husemann and W. Metzner,
{\it Incommensurate nematic fluctuations in the two-dimensional Hubbard model},
Phys. Rev. B {\bf 86}, 085113 (2012).

\bibitem{Kivelson-spin-nematic}
C. Fang, H. Yao, W.-F. Tsai, J. Hu, and S. A. Kivelson,
{\it Theory of electron nematic order in LaFeAsO},
Phys. Rev. B {\bf 77}, 224509 (2008).

\bibitem{DHLee-spin-nematic}
F. Wang, S. A. Kivelson, and D.-H. Lee,
{\it Nematicity and quantum paramagnetism in FeSe},
Nat. Phys. {\bf 11}, 959--963 (2015).

\bibitem{Tazai-CeB6}
R. Tazai and H. Kontani,
{\it Multipole fluctuation theory for heavy fermion systems: Application to multipole orders in ${\mathrm{CeB}}_{6}$},
Phys. Rev. B {\bf 100}, 241103(R) (2019).

\bibitem{Tazai-CeCu2Si2-1}
R. Tazai and H. Kontani,
{\it Fully gapped $s$-wave superconductivity enhanced by magnetic criticality in heavy-fermion systems},
Phys. Rev. B {\bf 98}, 205107 (2018).

\bibitem{Tazai-CeCu2Si2-2}
R. Tazai and H. Kontani,
{\it Hexadecapole Fluctuation Mechanism for $s$-wave Heavy Fermion Superconductor CeCu$_2$Si$_2$: Interplay between Intra- and Inter-Orbital Cooper Pairs},
J. Phys. Soc. Jpn. {\bf 88}, 063701 (2019).

\bibitem{mSR_TRS}
C. Mielke, D. Das, J.-X. Yin, H. Liu, R. Gupta, Y.-X. Jiang, M. Medarde, X. Wu, H. C. Lei, J. Chang, P. Dai, Q. Si, H. Miao, R. Thomale, T. Neupert, Y. Shi, R. Khasanov, M. Z. Hasan, H. Luetkens, and Z. Guguchia,
{\it Time-reversal symmetry-breaking charge order in a kagome superconductor},
Nature {\bf 602}, 245--250 (2022).

\bibitem{Kerr}
Q. Wu, Z. X. Wang, Q. M. Liu, R. S. Li, S. X. Xu, Q. W. Yin, C. S. Gong, Z. J. Tu, H. C. Lei, T. Dong, and N. L. Wang,
{\it Revealing the immediate formation of two-fold rotation symmetry in charge-density-wave state of Kagome superconductor CsV$_3$Sb$_5$ by optical polarization rotation measurement},
arXiv:2110.11306.

\bibitem{Murayama}
H. Murayama, K. Ishida, R. Kurihara, T. Ono, Y. Sato, Y. Kasahara, H. Watanabe, Y. Yanase, G. Cao, Y. Mizukami, T. Shibauchi, Y. Matsuda, and S. Kasahara,
{\it Bond Directional Anapole Order in a Spin-Orbit Coupled Mott Insulator ${\mathrm{Sr}}_{2}({\mathrm{Ir}}_{1\ensuremath{-}x}{\mathrm{Rh}}_{x}){\mathrm{O}}_{4}$},
Phys. Rev. X {\bf 11}, 011021 (2021).

\bibitem{Varma}
M. E. Simon and C. M. Varma,
{\it Detection and Implications of a Time-Reversal Breaking State in Underdoped Cuprates},
Phys. Rev. Lett. {\bf 89}, 247003 (2002).

\bibitem{Nersesyan}
A. A. Nersesyan, G. I. Japaridze, and I. G. Kimeridze,
{\it Low-temperature magnetic properties of a two-dimensional spin nematic state},
J. Phys.: Condens. Matter {\bf 3}, 3353 (1991).

\bibitem{Kivelson-NJP}
E. Berg, E. Fradkin, S. A. Kivelson, and J. M. Tranquada,
{\it Striped superconductors: how spin, charge and superconducting orders intertwine in the cuprates},
New J. Phys. {\bf 11}, 115004 (2009).

\bibitem{Halboth:2000tt}
C. J. Halboth and W. Metzner,
{\it Renormalization-group analysis of the two-dimensional Hubbard model},
Phys. Rev. B {\bf 61}, 7364 (2000).

\bibitem{Davis:2013ce}
J. C. S. Davis and D.-H. Lee,
{\it Concepts relating magnetic interactions, intertwined electronic orders, and strongly correlated superconductivity},
Proc. Natl. Acad. Sci. U.S.A. {\bf 110}, 17623 (2013).

\bibitem{Tazai-JPSJ-fRG}
R. Tazai, Y. Yamakawa, M. Tsuchiizu, and H. Kontani,
{\it $d$- and $p$-wave Quantum Liquid Crystal Orders in Cuprate Superconductors, $\ifmmode {\kappa}\else ${\kappa}$\fi{}$-(BEDT-TTF)$_2$X, and Coupled Chain Hubbard Models: Functional-renormalization-group Analysis},
J. Phys. Soc. Jpn. {\bf 90}, 111012 (2021).

\bibitem{Kontani-sLC}
H. Kontani, Y. Yamakawa, R. Tazai, and S. Onari,
{\it Odd-parity spin-loop-current order mediated by transverse spin fluctuations in cuprates and related electron systems},
Phys. Rev. Research {\bf 3}, 013127 (2021).

\bibitem{AGD}
A. A. Abrikosov, L. P. Gorkov, and I. E. Dzyaloshinski, {\it Methods of Quantum Field Theory in Statistical Physics}, (New York: Dover Publications, Inc., 1975.)

\bibitem{LW}
J. M. Luttinger and J. C. Ward,
{\it Ground-State Energy of a Many-Fermion System. II},
Phys. Rev. {\bf 118}, 1417 (1960).

\bibitem{Baym-Kadanoff}
G. Baym and L. P. Kadanoff,
{\it Conservation Laws and Correlation Functions},
Phys. Rev. {\bf 124}, 287 (1961).

\bibitem{Baym}
G. Baym,
{\it Self-Consistent Approximations in Many-Body Systems},
Phys. Rev. {\bf 127}, 1391 (1962).

\bibitem{FeSe-ARPES-Suzuki}
Y. Suzuki, T. Shimojima, T. Sonobe, A. Nakamura, M. Sakano, H. Tsuji, J. Omachi, K. Yoshioka, M. Kuwata-Gonokami, T. Watashige, R. Kobayashi, S. Kasahara, T. Shibauchi, Y. Matsuda, Y. Yamakawa, H. Kontani, and K. Ishizaka,
{\it Momentum-dependent sign inversion of orbital order in superconducting FeSe},
Phys. Rev. B {\bf 92}, 205117 (2015).

\bibitem{FeSe-Lif1}
M. Yi, H. Pfau, Y. Zhang, Y. He, H. Wu, T. Chen, Z. R. Ye, M. Hashimoto, R. Yu, Q. Si, D.-H. Lee, P. Dai, Z.-X. Shen, D. H. Lu, and R. J. Birgeneau,
{\it Nematic Energy Scale and the Missing Electron Pocket in FeSe},
Phys. Rev. X {\bf 9}, 041049 (2019).

\bibitem{FeSe-Lif2}
S. S. Huh, J. J. Seo, B. S. Kim, S. H. Cho, J. K. Jung, S. Kim, C. I. Kwon, J. S. Kim, Y. Y. Koh, W. S. Kyung, J. D. Denlinger, Y. H. Kim, B. N. Chae, N. D. Kim, Y. K. Kim, and C. Kim,
{\it Absence of Y-pocket in 1-Fe Brillouin zone and reversed orbital occupation imbalance in FeSe},
Commun. Phys. {\bf 3}, 52 (2020).

\bibitem{Eremin}
L. C. Rhodes, J. B\ifmmode \mbox{\"{o}}\else \"{o}\fi{}ker, M. A. M\ifmmode \mbox{\"{u}}\else \"{u}\fi{}ller, M. Eschrig, and I. M. Eremin,
{\it Non-local dxy nematicity and the missing electron pocket in FeSe},
npj Quantum Mater. {\bf 6}, 45 (2021).

\bibitem{Terashima}
T. Terashima, Y. Matsushita, H. Yamase, N. Kikugawa, H. Abe, M. Imai, S. Uji, S. Ishida, H. Eisaki, A. Iyo, K. Kihou, C.-H. Lee, T. Wang, and G. Mu,
{\it Elastoresistance measurements on ${\mathrm{CaKFe}}_{4}{\mathrm{As}}_{4}$ and ${\mathrm{KCa}}_{2}{\mathrm{Fe}}_{4}{\mathrm{As}}_{4}{\mathrm{F}}_{2}$ with the Fe site of ${C}_{2v}$ symmetry},
Phys. Rev. B {\bf 102}, 054511 (2020).

\bibitem{Mizukami-B2g}
Y. Mizukami, O. Tanaka, K. Ishida, M. Tsujii, T. Mitsui, S. Kitao, M. Kurokuzu, M. Seto, S. Ishida, A. Iyo, H. Eisaki, K. Hashimoto, and T. Shibauchi,
{\it Thermodynamic Signatures of Diagonal Nematicity in RbFe$_2$As$_2$ Superconductor},
arXiv:2108.13081.

\bibitem{Shibauchi-nemQCP}
K. Ishida, Y. Onishi, M. Tsujii, K. Mukasa, M. Qiu, M. Saito, Y. Sugimura, K. Matsuura, Y. Mizukami, K. Hashimoto, and T. Shibauchi,
{\it Pure nematic quantum critical point accompanied by a superconducting dome},
arXiv:2202.11674.

\bibitem{Zheng-NaFeAs}
C. G. Wang, Z. Li, J. Yang, L. Y. Xing, G. Y. Dai, X. C. Wang, C. Q. Jin, R. Zhou, and G.-q. Zheng,
{\it Electron Mass Enhancement near a Nematic Quantum Critical Point in ${\mathrm{NaFe}}_{1\ensuremath{-}x}{\mathrm{Co}}_{x}\mathrm{As}$},
Phys. Rev. Lett. {\bf 121}, 167004 (2018).

\bibitem{Onari-SCVC}
S. Onari and H. Kontani,
{\it Self-consistent Vertex Correction Analysis for Iron-based Superconductors: Mechanism of Coulomb Interaction-Driven Orbital Fluctuations},
Phys. Rev. Lett. {\bf 109}, 137001 (2012).

\bibitem{Schultz}
H. J. Schulz,
{\it Fermi-surface instabilities of a generalized two-dimensional Hubbard model},
Phys. Rev. B {\bf 39}, 2940(R) (1989).

\bibitem{Affleck}
I. Affleck and J. B. Marston,
{\it Large-n limit of the Heisenberg-Hubbard model: Implications for high-${T}_{c}$ superconductors},
Phys. Rev. B {\bf 37}, 3774(R) (1988).

\bibitem{Tazai-cLC}
R. Tazai, Y. Yamakawa, and H. Kontani,
{\it Emergence of charge loop current in the geometrically frustrated Hubbard model: A functional renormalization group study},
Phys. Rev. B {\bf 103}, L161112 (2021).

\bibitem{stoner}
E. P. Wohlfarth,
{\it Thermodynamic aspects of itinerant electron magnetism},
Physica B+C {\bf 91}, 305 (1977).

\bibitem{Potthoff}
M. Potthoff,
{\it Self-energy-functional approach to systems of correlated electrons},
Eur. Phys. J. B {\bf 32}, 429 (2003).

\bibitem{Yanase}
Y. Yanase and M. Ogata,
{\it Kinetic Energy, Condensation Energy, Optical Sum Rule and Pairing Mechanism in High-$T_c$ Cuprates},
J. Phys. Soc. Jpn. {\bf 74}, 1534 (2005).

\bibitem{Miyake}
T. Miyake, K. Nakamura, R. Arita, and M. Imada,
{\it Comparison of $Ab initio$ Low-Energy Models for LaFePO, LaFeAsO, BaFe$_2$As$_2$, LiFeAs, FeSe, and FeTe: Electron Correlation and Covalency},
J. Phys. Soc. Jpn. {\bf 79}, 044705 (2010).

\bibitem{Mermin}
H. Kontani and M. Ohno,
{\it Effect of a nonmagnetic impurity in a nearly antiferromagnetic Fermi liquid: Magnetic correlations and transport phenomena},
Phys. Rev. B {\bf 74}, 014406 (2006).

\bibitem{Chubukov-AL}
R.-Q. Xing, L. Classen, and A. V. Chubukov,
{\it Orbital order in FeSe: The case for vertex renormalization},
Phys. Rev. B {\bf 98}, 041108 (2018).

\bibitem{Kontani-ROP}
H. Kontani,
{\it Anomalous transport phenomena in Fermi liquids with strong magnetic fluctuations},
Rep. Prog. Phys. {\bf 71}, 026501 (2008).

\bibitem{Onari-smectic}
S. Onari and H. Kontani,
{\it Hidden antiferronematic order in Fe-based superconductor $\mathrm{Ba}{\mathrm{Fe}}_{2}{\mathrm{As}}_{2}$ and NaFeAs above ${T}_{S}$},
Phys. Rev. Research {\bf 2}, 042005(R) (2020).

\bibitem{FeSe-ARPES-Zhang}
Y. Zhang, M. Yi, Z.-K. Liu, W. Li, J. J. Lee, R. G. Moore, M. Hashimoto, M. Nakajima, H. Eisaki, S.-K. Mo, Z. Hussain, T. P. Devereaux, Z.-X. Shen, and D. H. Lu,
{\it Distinctive orbital anisotropy observed in the nematic state of a FeSe thin film},
Phys. Rev. B {\bf 94}, 115153 (2016).

\bibitem{Onari-B2g}
S. Onari and H. Kontani,
{\it Origin of diverse nematic orders in Fe-based superconductors: ${45}^{\ensuremath{\circ}}$ rotated nematicity in $A{\mathrm{Fe}}_{2}{\mathrm{As}}_{2}\phantom{\rule{4pt}{0ex}}(A=\text{Cs},\text{Rb})$},
Phys. Rev. B {\bf 100}, 020507(R) (2019).

\bibitem{Text-SCVC}
S. Onari and H. Kontani, { Iron-Based Superconductivity}, (ed. P.D. Johnson, G. Xu, and W.-G. Yin, Springer-Verlag Berlin and Heidelberg GmbH \& Co. K (2015)).

\bibitem{FeSe-specific-heat}
A. E. B\"ohmer, T. Arai, F. Hardy, T. Hattori, T. Iye, T. Wolf, H. v. L\"ohneysen, K. Ishida, and C. Meingast,
{\it Origin of the Tetragonal-to-Orthorhombic Phase Transition in FeSe: A Combined Thermodynamic and NMR Study of Nematicity},
Phys. Rev. Lett. {\bf 114}, 027001 (2015).

\bibitem{Kontani-Raman}
H. Kontani and Y. Yamakawa,
{\it Linear Response Theory for Shear Modulus ${C}_{66}$ and Raman Quadrupole Susceptibility: Evidence for Nematic Orbital Fluctuations in Fe-based Superconductors},
Phys. Rev. Lett. {\bf 113}, 047001 (2014).

\bibitem{saigo}
H. Kontani, T. Saito, and S. Onari,
{\it Origin of orthorhombic transition, magnetic transition, and shear-modulus softening in iron pnictide superconductors: Analysis based on the orbital fluctuations theory},
Phys. Rev. B {\bf 84}, 024528 (2011).

\bibitem{Onari-transport-FeSe}
S. Onari and H. Kontani,
{\it In-plane anisotropy of transport coefficients in electronic nematic states: Universal origin of nematicity in Fe-based superconductors},
Phys. Rev. B {\bf 96}, 094527 (2017).

\bibitem{Metzner-RMP}
W. Metzner, M. Salmhofer, C. Honerkamp, V. Meden, and K. Sch\"onhammer,
{\it Functional renormalization group approach to correlated fermion systems},
Rev. Mod. Phys. {\bf 84}, 299 (2012).

\bibitem{Honerkamp}
M. Salmhofer and C. Honerkamp,
{\it Fermionic Renormalization Group Flows: Technique and Theory},
Prog. Theor. Phys. {\bf 105}, 1 (2001).

\bibitem{Tsuchiizu-PRL2013}
M. Tsuchiizu, Y. Ohno, S. Onari, and H. Kontani,
{\it Orbital Nematic Instability in the Two-Orbital Hubbard Model: Renormalization-Group + Constrained RPA Analysis},
Phys. Rev. Lett. {\bf 111}, 057003 (2013).

\bibitem{Kontani-PRL2010}
H. Kontani and S. Onari,
{\it Orbital-Fluctuation-Mediated Superconductivity in Iron Pnictides: Analysis of the Five-Orbital Hubbard-Holstein Model},
Phys. Rev. Lett. {\bf 104}, 157001 (2010).

\bibitem{Yamakawa-FeSe-underP}
Y. Yamakawa and H. Kontani,
{\it Nematicity, magnetism, and superconductivity in FeSe under pressure: Unified explanation based on the self-consistent vertex correction theory},
Phys. Rev. B {\bf 96}, 144509 (2017).

\bibitem{Yamakawa-FeSe-SC}
Y. Yamakawa and H. Kontani,
{\it Superconductivity without a hole pocket in electron-doped FeSe: Analysis beyond the Migdal-Eliashberg formalism},
Phys. Rev. B {\bf 96}, 045130 (2017).

\bibitem{Bickers}
N. E. Bickers, D. J. Scalapino, and S. R. White,
{\it Conserving Approximations for Strongly Correlated Electron Systems: Bethe-Salpeter Equation and Dynamics for the Two-Dimensional Hubbard Model},
Phys. Rev. Lett. {\bf 62}, 961 (1989).

\bibitem{Onari-SCVCS}
S. Onari, Y. Yamakawa, and H. Kontani,
{\it High-${T}_{c}$ Superconductivity near the Anion Height Instability in Fe-Based Superconductors: Analysis of ${\mathrm{LaFeAsO}}_{1\ensuremath{-}x}{\mathrm{H}}_{x}$},
Phys. Rev. Lett. {\bf 112}, 187001 (2014).

\bibitem{FeSe-t}
A. E. B\"ohmer, F. Hardy, F. Eilers, D. Ernst, P. Adelmann, P. Schweiss, T. Wolf, and C. Meingast,
{\it Lack of coupling between superconductivity and orthorhombic distortion in stoichiometric single-crystalline FeSe},
Phys. Rev. B {\bf 87}, 180505(R) (2013).

\end{thebibliography}
\end{document}